\documentclass[]{aa} 
\usepackage{amssymb}
\usepackage{hyperref}
\usepackage{graphicx}
\usepackage{txfonts}
\def\asec{\ifmmode ^{\prime\prime}\else$^{\prime\prime}$\fi}

\def\msun{M$_{\odot}$}

\def\mdot{$\dot M$}

\def\degs{\ifmmode ^{\circ}\else$^{\circ}$\fi}
\def\amin{\ifmmode ^{\prime}\else$^{\prime}$\fi}
\def\asec{\ifmmode ^{\prime\prime}\else$^{\prime\prime}$\fi}
\def\farcs{\hbox{$.\!\!^{\prime\prime}$}}  
\def\lsim{\!\!\!\phantom{\le}\smash{\buildrel{}\over
 {\lower2.5dd\hbox{$\buildrel{\lower2dd\hbox{$\displaystyle<$}}\over
                               \sim$}}}\,\,}
\def\gsim{\!\!\!\phantom{\ge}\smash{\buildrel{}\over
 {\lower2.5dd\hbox{$\buildrel{\lower2dd\hbox{$\displaystyle>$}}\over
                               \sim$}}}\,\,}

\def\Msunyr{~{\rm M}_\odot~{\rm yr}^{-1}}
\def\EE#1{\times 10^{#1}}
\def\kms{\rm ~km~s$^{-1}$}

\begin{document}
\title{Early and late time VLT spectroscopy of SN~2001el $-$ progenitor
constraints for a type Ia supernova\thanks{Based on observations collected at 
the European Southern Observatory, Paranal, Chile (ESO Programmes 67.D-0227 
and 69.D-0193). Included in the `The Physics of Type Ia SNe' EC Programme (HPRN-CT-2002-00303; 
PI: W.~Hillebrandt)}}

\author{S.~Mattila\inst{1}
\and P.~Lundqvist\inst{1}
\and J.~Sollerman\inst{1}
\and C.~Kozma\inst{1}
\and E.~Baron\inst{2}
\and C.~Fransson\inst{1}
\and B.~Leibundgut\inst{3}
\and K.~Nomoto\inst{4}
          }
\authorrunning{Mattila et al.}
\titlerunning{Early and late time VLT spectroscopy of SN~2001el} 
   \offprints{S. Mattila;\hfill\\
e-mail: seppo@astro.su.se
}

\institute{ Stockholm Observatory, AlbaNova, Department of Astronomy,
SE-106 91 Stockholm, Sweden 
\and Department of Physics and Astronomy, University of Oklahoma, 440
West Brooks Street, Norman, OK 73019, U.S.A.
\and European Southern Observatory, Karl-Schwarzschild-Strasse 2, 
D-85748 Garching bei M\"unchen, Germany
\and Department of Astronomy and Research Center for the Early Universe, 
University of Tokyo, Bunkyo-ku, Tokyo, Japan
             }

   \date{Received ??; accepted ??}

   \abstract{
We present early time high-resolution (VLT/UVES) and late time 
low-resolution (VLT/FORS) optical spectra of the normal type Ia 
supernova, SN~2001el. The high-resolution spectra were obtained 9 and 2 days before 
(B-band) maximum light. This was in order to allow the detection of narrow 
hydrogen and/or helium emission lines from the circumstellar medium of the 
supernova. No such lines were detected in our data. We therefore use these spectra together 
with photoionisation models to derive upper limits of $9\times10^{-6} \Msunyr$ 
and $5\times10^{-5} \Msunyr$ for the mass loss rate from the progenitor system of SN~2001el
assuming velocities of 10 \kms~and 50 \kms, respectively, for a wind extending to outside 
at least a few $\times 10^{15}$ cm away from the supernova explosion site. 
So far, these are the best H$\alpha$ based upper limits obtained for a type Ia 
supernova, and exclude a symbiotic star in the upper mass loss rate regime (so 
called Mira type stars) from being the progenitor of SN~2001el. The 
low-resolution spectrum was obtained in the nebular phase of the supernova, 
$\sim$400 days after the maximum light, to search for any hydrogen rich gas 
originating from the supernova progenitor system. However, we see no signs of 
Balmer lines in our spectrum. Therefore, we model the late time spectra to 
derive an upper limit of $\sim$0.03 M$_{\odot}$ for solar abundance material 
present at velocities lower than 1000 \kms~within the supernova explosion 
site. According to numerical simulations of Marietta et al. (2000) this is 
less than the expected mass lost by a subgiant, red giant or a main-sequence 
secondary star at a small binary separation as a result of the SN explosion. 
Our data therefore exclude these scenarios as the progenitor of SN 2001el. 
Finally, we discuss the origin of high velocity Ca~II lines previously 
observed in a few type Ia supernovae before the maximum light. We see both 
the Ca~II IR triplet and the H$\&$K lines in our earliest ($-$9 days) spectrum 
at a very high velocity of up to $\sim$34~000 km~s$^{-1}$. The spectrum also 
shows a flat-bottomed Si~II `6150 \AA' feature similar to the one previously 
observed in SN~1990N (Leibundgut et al. 1991) at 14 days before maximum light. 
We compare these spectral features in SN~2001el to those observed in SN~1984A 
and SN~1990N at even higher velocities. 

   }

   \maketitle
%

\section{Introduction}

While the progenitor stars for a few core-collapse supernovae (SNe), viz. SN~1987A
(White \& Malin 1987; Walborn et al. 1989), SN~1993J (Aldering et al. 1994; 
Maund et al. 2004) and SN~2003gd (Smartt et al. 2004), have
been directly detected in pre-explosion images, and a general good picture
has emerged from stellar evolutionary modelling (Smartt et al. 2003; Eldridge
\& Tout 2004), similar constraints are still missing for the progenitors of 
thermonuclear (type Ia) supernovae (henceforth SNe Ia). SNe Ia are believed 
to arise from the thermonuclear explosion of a carbon oxygen white dwarf 
(CO WD) in a binary system when its degenerate mass becomes close to the 
Chandrasekhar limit. Models predict that the companion star can either be 
another CO WD or a non-degenerate star. In the double degenerate scenario two 
low-mass CO WDs merge, whereas in the single degenerate scenario matter is 
accreted from the 
companion star until the CO WD reaches its limiting mass for carbon ignition (see
Hillebrandt \& Niemeyer 2000 and Nomoto et al. 2000 for discussion and references). In neither scenario would the 
progenitor system be luminous enough to be directly detectable 
at extragalactic distances. However, the recently reported (Ruiz-Lapuente et al. 2004)
likely identification of the subgiant/main-sequence companion star surviving the explosion 
of Tycho Brahe's supernova (SN~1572) gives support for the single degenerate scenarios.

In all the single degenerate scenarios hydrogen and/or helium rich gas 
originating from the companion star could be present in the circumstellar medium 
(CSM) of the SN. This gas could be in the form of a stellar wind, an accretion
disk, a filled Roche-lobe, a common envelope around the progenitor system, or 
a wind from the accreting WD (Hachisu et al. 1999a, 1999b).
This material could be detectable by making use of the SN as a probe of its own 
CSM. If the wind is dense enough and ionised by the radiation from the SN 
ejecta/wind interaction, narrow emission lines similar to those in narrow-line 
(type IIn) core-collapse SNe (and SN 1987A at late times) could be visible on 
top of the SN spectrum (Cumming et al. 1996; Lundqvist et al. 2006). In fact, 
SN~2002ic was recently observed to show spectral features similar to both SNe 
Ia and SNe IIn (Hamuy et al. 2003; Baron 2003; Deng et al. 2004; Kotak et al. 
2004); it could belong to the so-called type 1.5 SNe (e.g., Chugai et al. 
2004; Nomoto et al. 2004) which were predicted by Iben \& Renzini (1983). 
The SN~Ia explosion itself is also expected to cause a loss of material 
from a close, non-merging companion (Fryxell \& Arnett 1981; Taam \& 
Fryxell 1984; Livne et al. 1992; Marietta et al. 2000). This material is 
likely to be hydrogen rich, and could reveal itself when the ejecta have 
become transparent.

In addition, Ca~II IR triplet (8498~\AA, 8542~\AA, 8662~\AA) and H\&K 
(3968~\AA, 3934~\AA) lines have recently been observed at very high velocities 
compared to the SN photospheric lines in a few SNe Ia: SN~1994D (Hatano et al. 
1999), SN~1999ee (Mazzali et al. 2005a), SN~2000cx (Thomas et al. 2004; Branch 
et al. 2004), 
SN~2001el (Wang et al. 2003; Kasen et al. 2003), SN~2003du (Gerardy et al. 2004) and 
SN~2004dt (Wang et al. 2004). While the origin of these high velocity (HV) lines 
is still unknown, one of the interpretations has been interaction of the SN 
ejecta with CSM material originating from the companion star (e.g., Gerardy et al.
2004).

To shed light on the origin of SNe Ia we have been conducting a Target of 
Opportunity (ToO) programme on the European Southern Observatory (ESO) Very Large 
Telescope (VLT). We have observed two SNe Ia so far, SNe 2000cx and 
2001el with the VLT (for preliminary analyses see Lundqvist et al. 2003; 
Lundqvist et al. 2005). The observations of SN~2000cx, and their modelling, 
are described in Lundqvist et al. (2006). Here we report our observations of 
SN~2001el.

Supernova 2001el was discovered on September 17.1 UT (Monard 2001) 
about 14" west and 20" north of the nucleus of an edge-on spiral galaxy
NGC~1448. This nearby ($V_{\rm rec}$ = 1164 km~s$^{-1}$; de Vaucouleur et al. 
1991) spiral has hosted also two type II SNe (SN~1983S and SN~2003hn) during
the last 20 years. The magnitude of SN~2001el at the time of the discovery was
about 14.5. It was classified as a SN Ia by us on September 21.3 UT (Sollerman et
al. 2001). SN~2001el was observed to show a polarised spectrum before the 
maximum light (Wang et al. 2003) indicating some asymmetric structure 
during these initial phases of expansion. Later on, SN~2001el has been 
well monitored both photometrically and spectroscopically, and has been shown 
to be a normal SN Ia with $\Delta$m$_{15}$(B)=1.13 (Krisciunas et al. 2003). 

In Sect.~2 the VLT observations and data reductions are described. In 
Sect.~3, we describe how we searched for CSM lines from the early time
high-resolution and late time low-resolution spectra of SN~2001el. No CSM lines 
were detected and thus upper limits were derived for the line fluxes. In Sect.~4, 
we use these flux limits together with models to estimate upper limits for the amount 
of hydrogen rich material lost from the companion star either via a stellar wind or 
as a result of the SN explosion. In Sect.~5, we compare the profiles of the high velocity
Ca~II IR triplet and H$\&$K lines as well as the flat-bottomed Si~II `6150 \AA' feature 
in our early time spectrum to those observed in SNe 1984A and 1990N at a similar
early epoch. Finally, in Sect.~6 we discuss our results and present the summary.


\begin{table*}[t]
\caption{Log of VLT observations of SN~2001el.}
\begin{tabular}{lllllllll}
\hline
\hline
Instrument & Date (UT)     & JD       & Epoch & Exp.  & a.m.$^a$ & Seeing$^b$ & set-up & Slit width\\ 
           &               &(2452000+)& (days)& (sec) &          & (arcsec)   &        & (arcsec) \\
\hline

UVES       & 2001 Sep 21.22 & 173.72       & $-8.7$    & 2400 & 1.30 & 0.8 & 390+564$^c$ & 0.8 \\
UVES       & 2001 Sep 21.25 & 173.75       & $-8.7$    & 2400 & 1.19 & 0.8 & 390+564 & 0.8 \\
UVES       & 2001 Sep 21.28 & 173.78       & $-8.7$    & 2400 & 1.12 & 0.7 & 437+860$^d$ & 0.8  \\
UVES       & 2001 Sep 21.31 & 173.81       & $-8.7$    & 2400 & 1.08 & 0.9 & 437+860 & 0.8 \\

UVES       & 2001 Sep 28.22 & 180.72       & $-1.7$    & 3000 & 1.20 & 1.1 & 390+564 & 0.8 \\
UVES       & 2001 Sep 28.26 & 180.76       & $-1.7$    & 3000 & 1.11 & 1.1 & 390+564 & 0.8 \\
UVES       & 2001 Sep 28.29 & 180.79       & $-1.7$    & 3000 & 1.07 & 1.1 & 390+564 & 0.8 \\

\hline

FORS1      & 2002 Nov 02.15  & 580.65        & $+398$ & 3000   & 1.15 & 1.2  & GRIS$\_$300V & 1.3 \\

\hline
\end{tabular} 
\\

\begin{tabular}{lll}
$^a$ \ The effective airmass of the observation \\
$^b$ \ Seeing from the DIMM-monitor \\
$^c$ \ Setting 390+564 covers wavelength ranges 3260$-$4450~\AA, 4580$-$6680~\AA. \\
$^d$ \ Setting 437+860 covers wavelength range 3730$-$4990~\AA, 6600$-$10600~\AA. \\
\end{tabular}
\label{t:obs}
\end{table*}

\section{VLT observations and data reductions}

\subsection{High-resolution UVES spectroscopy before the maximum light}
Two days after the discovery IAU circular (Monard 2001) on September 21.3 (UT),
SN~2001el, still without a spectroscopic classification, was observed with the 
Ultraviolet and Visual Echelle Spectrograph (UVES) on the VLT. All the
observations were obtained in service mode. We obtained 
two 2400 second exposures in each of two different dichroic settings, DIC1 
390+564 and DIC2 437+860 (Table 1). These settings together give a complete wavelength 
coverage between 3260~\AA~and 10600~\AA. The night was clear and the seeing 
during the observations was around 0\farcs8. The spectrophotometric standard 
star LTT~1020 (Hamuy et al. 1994) was also observed in both setups, after the supernova 
observations. All spectra were quickly reduced using the UVES pipeline.
This allowed a classification of the supernova as a type Ia, well before the maximum 
light (Sollerman et al. 2001). 

The SN was observed again, 7 days later, on September 28.3 (UT) 
to enable monitoring of any weak CSM lines potentially detectable from this 
promising early SN Ia event. Three exposures of 3000 seconds each 
were obtained using the dichroic setting DIC1 390+564, which covers 
the wavelength ranges 3260$-$4450~\AA~and 4580$-$6680~\AA. The night was clear 
and the seeing was around 1\farcs1 during the observations. The 
spectrophotometric standard star HR~1996 (Tinney \& Reid 1998) was observed immediately after 
the supernova observations. The 0\farcs8 slit was used for all the observations 
on both epochs yielding a spectral resolution of R $\sim$ 50~000 ($\sim$6 \kms)
for the region around H$\alpha$. Our UVES observations of SN~2001el were thus obtained 8.7 and 
1.7 days before the supernova B-band maximum light, which occurred on September 
30.0 (UT) (Krisciunas et al. 2003).

\begin{figure*}[t]
\begin{minipage}{180mm}
\includegraphics[width=140mm, angle=-90, clip] 
{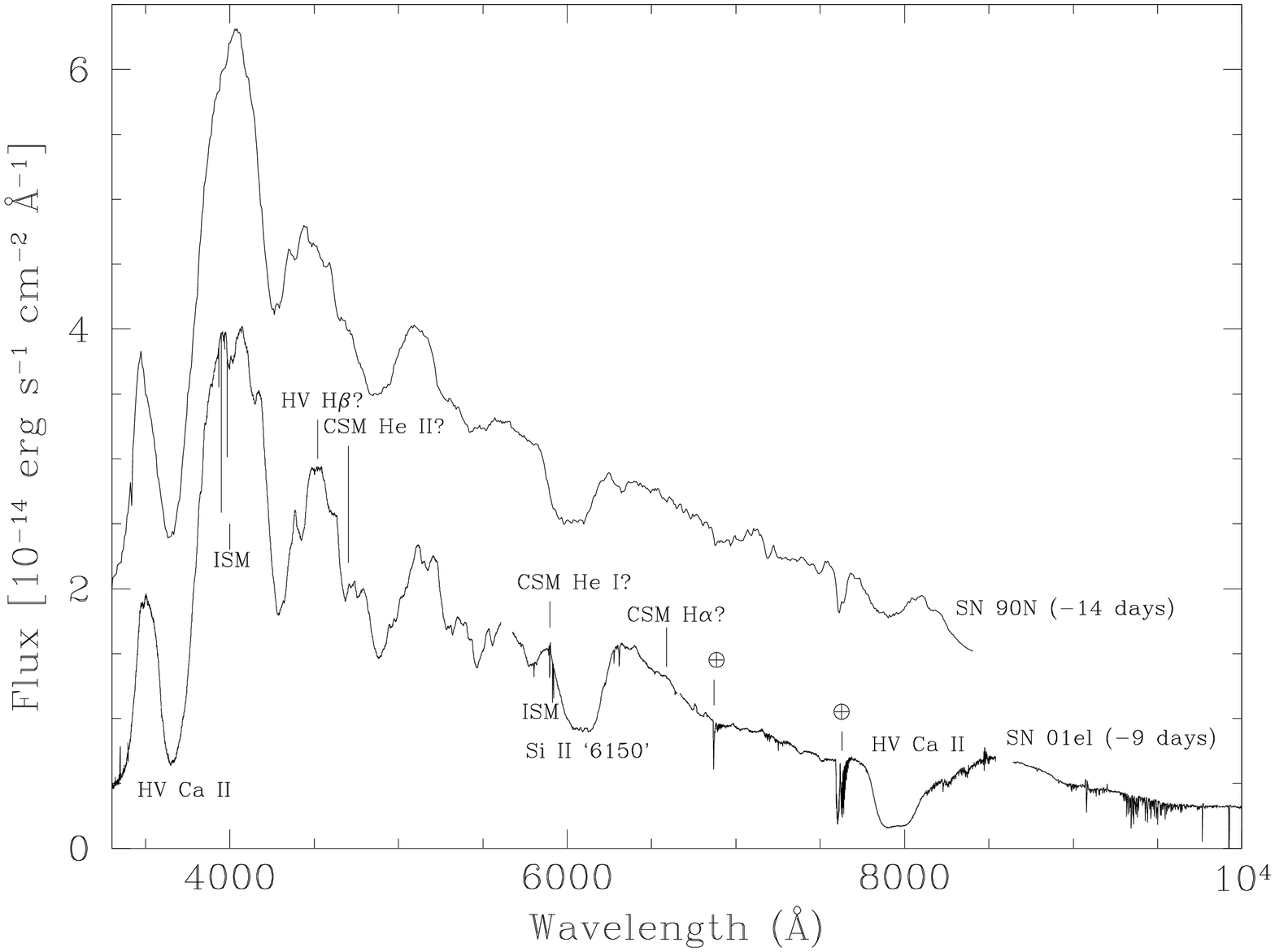}
\caption{
UVES spectrum of SN~2001el obtained on September 21.3 (UT), i.e.,
8.7 days before the B-band maximum light. The spectrum has been median filtered
and rebinned to have a resolution of about 50 km~s$^{-1}$~pixel$^{-1}$.
The lines of interest for this study are indicated. The spectrum of SN~1990N observed (Leibundgut et al. 1991)
at 14 days before the B-band maximum light is also shown for comparison. This spectrum has been redshifted 
to match the recession velocity of SN~2001el, scaled by $\times$3.5, and shifted upward by 
1.5$\times$10$^{-14}$ erg~s$^{-1}$~cm$^{-2}$\AA$^{-1}$ for clarity. The Ca~II IR triplet and the
H$\&$K lines are present at a very high velocity (HV), and the characteristic Si~II
`6150 \AA' absorption shows a similar strongly blueshifted flat-bottomed profile in both SNe. 
However, we see no signs of narrow circumstellar hydrogen or helium lines nor HV H$\beta$ in the SN~2001el 
spectrum. The Ca~II and Na~I ISM lines are also marked in the SN~2001el spectrum (for details see Sollerman
et al. 2005). A large number of telluric lines are present, especially at wavelengths
longer than 9000 \AA.}
\label{f:2}
\end{minipage} 
\end{figure*}

\subsection{UVES data reductions}
The spectral orders covering the lines of interest for this study were carefully
reduced using a combination of tasks from the UVES pipeline\footnote
{http://www.eso.org/observing/dfo/quality/} (Ballester et al. 2003)
(version 2.0) and IRAF.
All the 2D data were interactively reduced using the UVES-pipeline
task $\tt REDUCE/SPAT$ and calibration frames obtained during daytime 
operation. This included bias subtraction, flat fielding the data in pixel space,
automatically determining the Echelle order positions and slopes, and 
subtraction of the inter-order (median) background. The pipeline performed 
wavelength calibration for the 2D data by automatically identifying and fitting 
a large number ($\sim$ 1300 per set-up) of ThAr arc lines. For the 
wavelength calibration a single dispersion solution, obtained for the center coordinate 
of the object, is used for each pixel in the spatial direction. The pipeline then 
produces separate 2D frames for each Echelle order with constant wavelength binning on the x-axis, 
and the spatial extent of the data on the y-axis.

An accurate correction for the background emission from the supernova host 
galaxy is vital when searching for traces of weak emission or absorption lines from the 
supernova CSM (see also Della Valle et al. 1996). In fact, the widths 
and luminosities of emission lines expected from the CSM of a SN Ia
are similar to the ones observed from ordinary H~II regions. The luminosity 
time evolution of such circumstellar emission lines can also be slow 
enough (in a timespan of a few days) to result in confusion with the constant 
emission from a nearby H~II region. We performed the background subtraction outside 
the UVES pipeline using the $\tt IRAF~BACKGROUND$ task interactively. 
We found that the dispersion solution obtained for the center coordinate of the object 
was valid over the slit length (8$-$12$\arcsec$) enabling accurate subtraction of the 
sky and host galaxy background emission from the wavelength calibrated frames.

The UVES pipeline uses a Gaussian approximation for the cross-order light 
profile during the optimal extraction. This is not valid when the 
signal to noise ratio (S/N) is high enough (Piskunov \& Valenti 2002;
Ballester et al. 2003). In fact, we found that the standard optimal extraction 
routines included, e.g., in the UVES pipeline and $\tt IRAF$ did not give satisfactory 
results for our Echelle spectra with high S/N. Therefore, in this study we decided to use  
simple average extraction for the SN and standard star spectra, and performed 
the cosmic ray rejection when combining the 1D spectra. 

To correct for the telluric absorption lines and any residual instrumental features 
left after the flat fielding, the supernova spectra were divided by the spectrum of the 
standard star observed during the same night as the supernova with the same instrument settings 
and scaled to match the effective airmass of the science observation. Both the standard stars 
observed, LTT~1020 and HR~1996, show broad H$\alpha$ absorption lines with their wings extending 
close to the region of interest when searching for the circumstellar H$\alpha$ line from the 
supernova spectrum. We therefore normalised the wavelength regions of interest in the standard star spectra 
with a low order polynomial fit before the division of the supernova data. The LTT~1020 spectrum was 
also smoothed before the division in order to conserve the S/N of the science data. The corrected science 
exposures were then normalised by a low order polynomial fit, and average combined together (see Figs. 3-4). For 
this we used weighting by the inverse squares of the corresponding pixel to pixel noise levels. 
Any deviant pixels or cosmic rays (above 5$\sigma$ level) were rejected and replaced by the average of the two 
unaffected spectra or in the case of only two available spectra by a fit to the continuum. The 
individual orders covering the wavelengths around the Ca~II (H\&K and the IR triplet) and the Si~II 
`6150 \AA' lines were carefully merged together in {\tt IRAF} (see Fig.~8).

The absolute flux calibration was performed for each wavelength region
of interest separately. For this we used the well-sampled V-band light 
curve of SN~2001el from Krisciunas et al. (2003) and a low-resolution spectrum
from Wang et al. (2003). This spectrum was observed on September 26.3 (UT),
i.e., 4.0 days before the maximum light. To estimate the uncertainties 
due to the spectral evolution between the epochs of our UVES spectra and 
the spectrum of Wang et al. (2003) we used low-resolution spectra of SN~2002bo 
(Benetti et al. 2004) from 9.0, 4.0 and 2.0 days before the maximum light.
These are both normal SNe Ia with exactly the same $\Delta$m$_{15}$ values
(Krisciunas et al. 2003; Benetti et al. 2004). The 
low-resolution spectra were scaled in flux to match the observed 
$V$-band photometry at the dates of our UVES observations. From this comparison 
we estimate that our absolute flux calibration is accurate to better than $\pm$25$\%$.

The reduction procedure outlined above was designed to mainly study
limited wavelength regions around narrow, and presumably weak line
features. To obtain an overall spectrum of the SN we do not need such a high
accuracy. For the first epoch ($-8.7$ day) overall spectrum we therefore
simply used the standard UVES pipeline task $\tt REDUCE/UVES$. Here the individual orders 
were automatically extracted and merged 
together by the pipeline. To enable successful automatic merging, simple average extraction 
was used, and the individual orders were flat fielded in 1D. The relative flux calibration 
was carried out in $\tt IRAF$ separately for the spectra corresponding to each spectrograph set-up 
(see Table 1). For this we used the standard star (LTT~1020) spectra observed
together with the SN. To remove the large number of cosmic rays we median filtered 
each individual spectrum using a filter window of 50 pixels ($\sim$50 km~s$^{-1}$). 
The overlapping spectra were average combined together, and rebinned to yield a pixel scale 
of $\sim$50 km~s$^{-1}$~pixel$^{-1}$. The resulting spectrum is shown in Fig.~1 where the levels 
of the wavelength regions covered by the different set-ups have been adjusted by eye to match the 
overall shape of the $-4$ day low-resolution spectrum of Wang et al. (2003). The overall level of the 
spectrum has been scaled in flux to match the observed $V$-band photometry of the SN. 

\subsection{Low-resolution FORS spectroscopy in the nebular phase}
We have also obtained a very late low-resolution optical spectrum of SN~2001el.
On November 2.2 (UT) 2002, 398 days past maximum light, we used FORS1 on the VLT
to obtain a 3000 second exposure (see Table~1 for details). 
We used the 300V grism together with order-sorting filter GG375 and 
a $1\farcs3$ wide slit yielding a spectral resolution of R $\sim$ 440 ($\sim$700 \kms)
for the observation.

\begin{figure}[t]
\begin{minipage}{85mm}
\includegraphics[width=80mm, angle=0, clip] 
{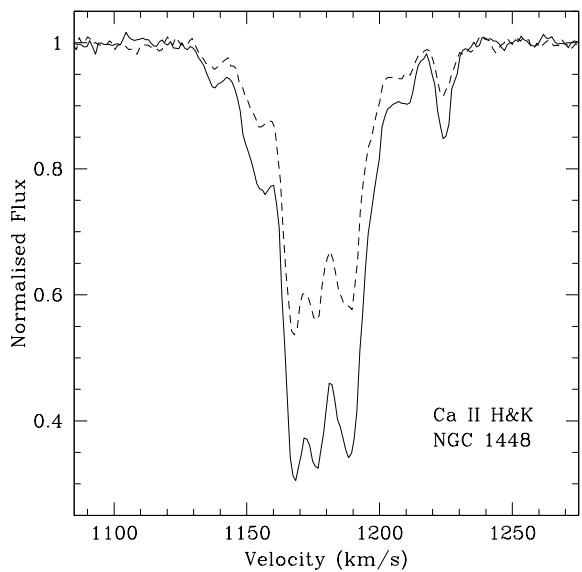}
\caption{Ca~II H ($\lambda$3968, dashed) and K ($\lambda$3934, solid) 
interstellar absorption lines towards SN~2001el 
within its host galaxy NGC~1448. The velocity scale is heliocentric. See 
Sollerman et al. (2005) for details about the absorption components.}
\end{minipage} 
\label{f:1}
\end{figure}

The spectrum (see Fig.~6) was reduced in a standard way in $\tt IRAF$, including bias
subtraction, flat fielding, and wavelength calibration using spectra of
a Helium-Argon lamp.  Flux calibration was done relative to the
spectrophotometric standard star LTT~2415 (Hamuy et al. 1994).
We have also compared the synthetic $V$ band magnitude obtained from
this spectrum to late time light curves of SNe Ia
(e.g., Sollerman et al. 2004, their Fig.~11) and conclude that
the absolute fluxing of the spectrum is accurate to $\pm$20\%.

\section{Searching for traces of circumstellar lines}

\subsection{The expected wavelength range for the CSM lines}
Superposed on the supernova spectra (Fig.~1) we also detect interstellar 
Ca~II H\&K absorption both from the Galaxy and from NGC~1448 which hosted the supernova. 
The Ca~II H\&K absorption lines arising within the host galaxy are plotted in Fig.~2
as observed on Sept. 28. Several absorption components are present in the spectrum (see detailed analysis
in Sollerman et al. 2005). The overall range of velocities for the absorbing 
gas along the line-of-sight to the supernova 
is $V_{\rm abs}$ $\sim$ 1130$-$1230 km~s$^{-1}$. The recession velocity of the 
H II regions close to the supernova included in the UVES slit can be used as another indicator of the 
supernova recession velocity. We measured $V_{\rm H~II}$ = 
(1186 $\pm$ 45) km~s$^{-1}$ for the H$\alpha$ emitting gas next to the SN
position in the UVES spectrum at Sept. 28. 
As NGC~1448 is oriented almost edge on ($i$ = 88$^{\circ}$) the galaxy rotation 
may result in a substantial velocity for the supernova w.r.t. to the host 
galaxy recession velocity. A rotation velocity of $V_{\rm rot}$ $\sim$193 
km~s$^{-1}$ has been measured for NGC~1448 (Mathewson $\&$ Ford 1996). 
We therefore carry-out the search for the narrow circumstellar emission lines 
at the range of velocities $V_{\rm rec}$ = (1180 $\pm$ 190) km~s$^{-1}$. 
This is close to the recession velocity of NGC~1448 of 1164 $\pm§$ 5 km~s$^{-1}$
compiled by de Vaucouleur et al. (1991).  

\begin{figure*}
\begin{center}
\includegraphics[width=80mm, angle=-90, clip] {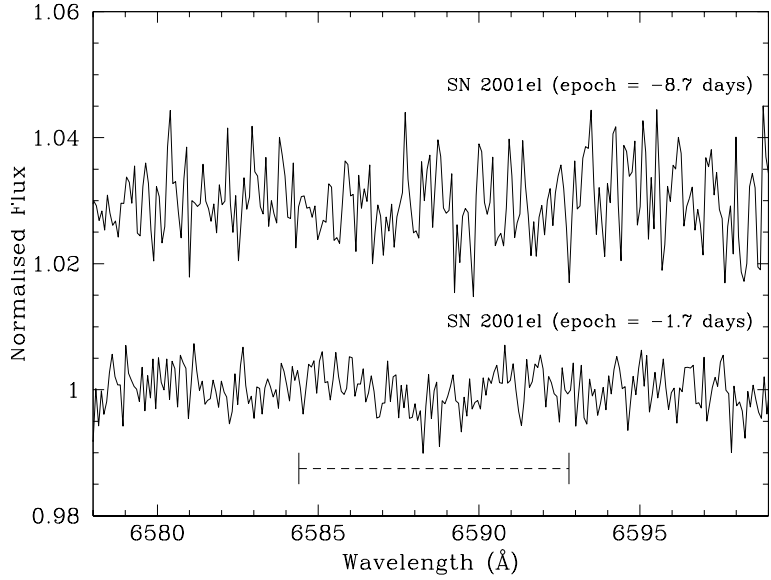}
\end{center}  
\caption{Normalised and rebinned ($\sim$4 km~s$^{-1}$~pixel$^{-1}$) UVES spectra 
in the expected spectral region around 
H$\alpha$ for SN~2001el on two epochs, September 21.3 and 28.3 (UT) 2001, 
i.e., 8.7 and 1.7 days before the SN maximum light, respectively. The expected 
wavelength range of H$\alpha$ is marked with a horizontal dashed line, and
the upper spectrum has been shifted vertically for clarity. 
No significant emission or absorption lines are visible.}
\label{f:3}
\end{figure*}

\begin{figure*}
\begin{center}
\includegraphics[width=60mm, angle=-90, clip] {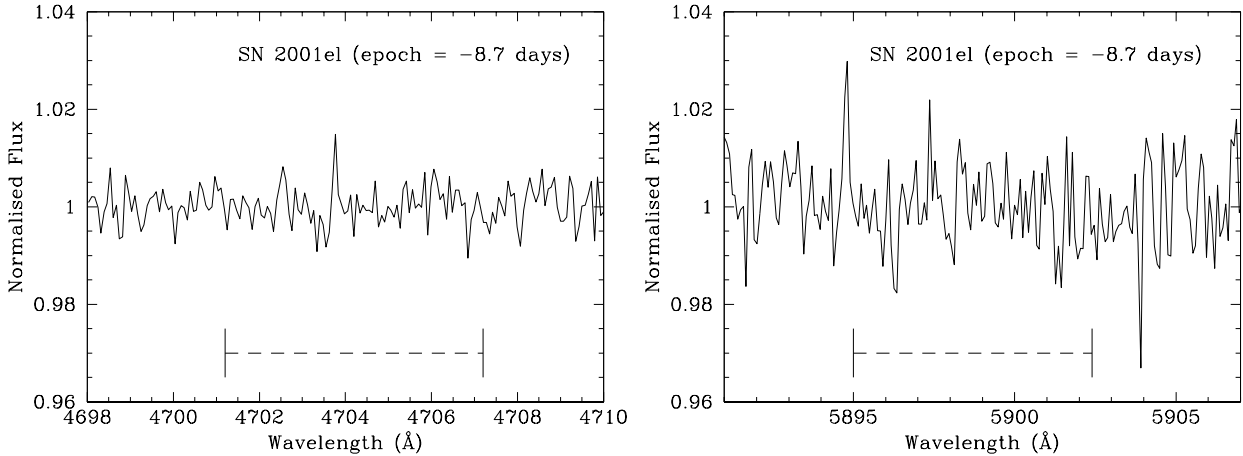}
\end{center}  
\caption{Normalised and rebinned ($\sim$4 km~s$^{-1}$~pixel$^{-1}$) UVES spectra in the expected 
spectral regions around He~II
(4686~\AA) and He~I (5876~\AA) for SN~2001el on September 21.3 (UT) 2001, i.e.,
8.7 days before the maximum light. The expected wavelength range of the CSM 
lines is marked with a horizontal dashed line. No significant emission or absorption lines are visible.
}
\label{f:4}
\end{figure*}

\begin{figure*}[t]
\begin{center}
\includegraphics[width=60mm, angle=0, clip] {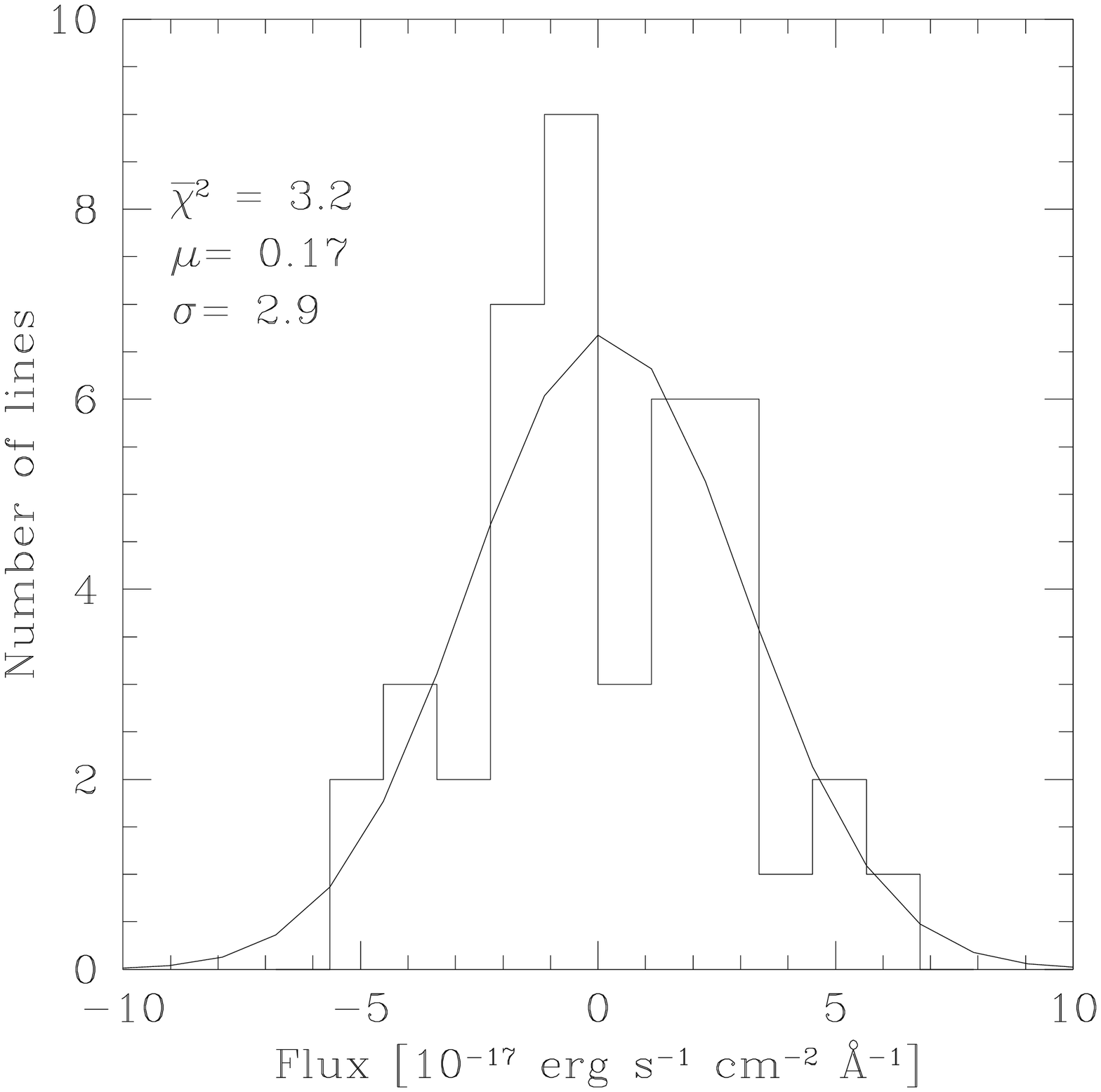}
\includegraphics[width=60mm, angle=0, clip] {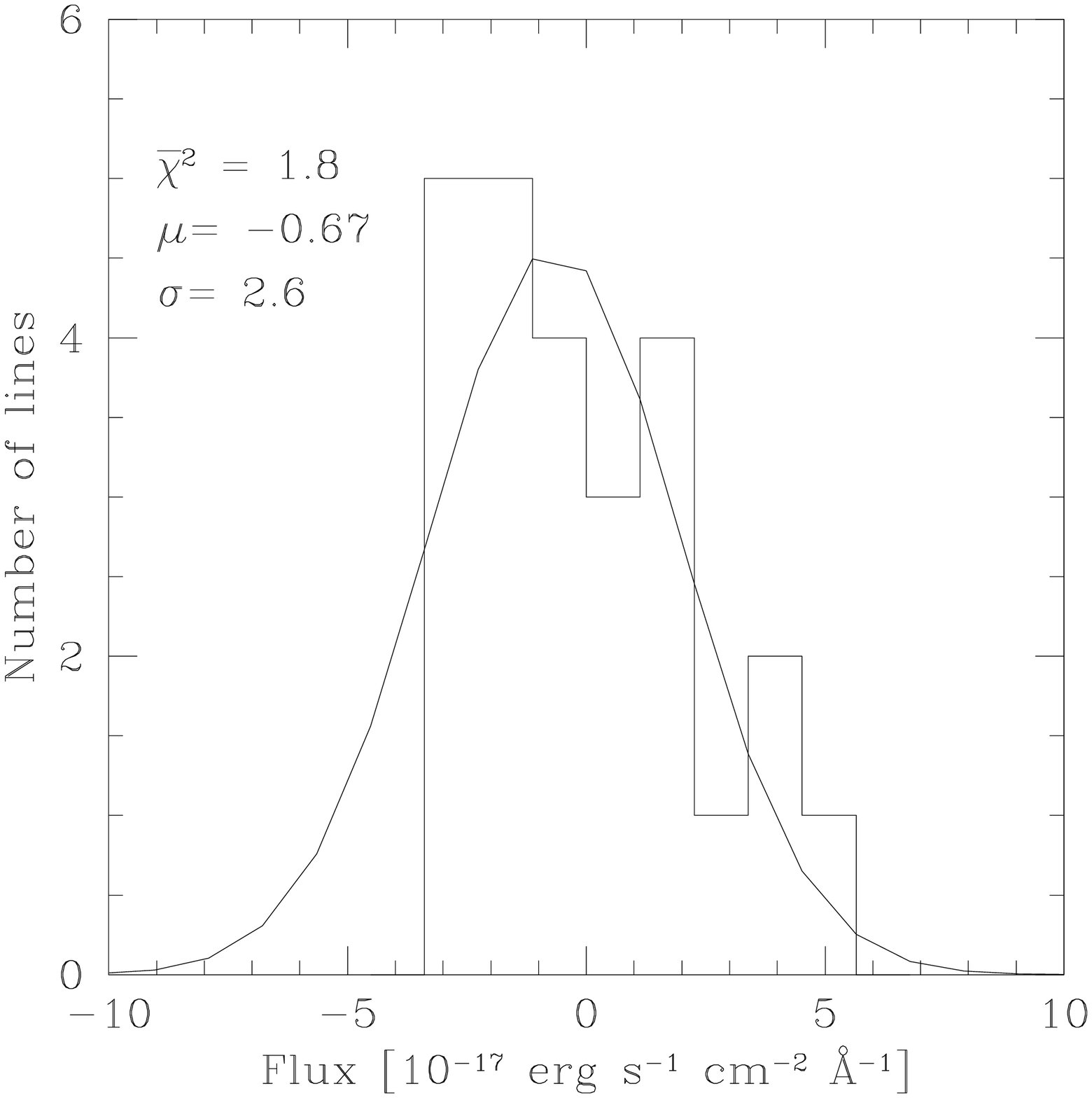}
\includegraphics[width=60mm, angle=-90, clip]{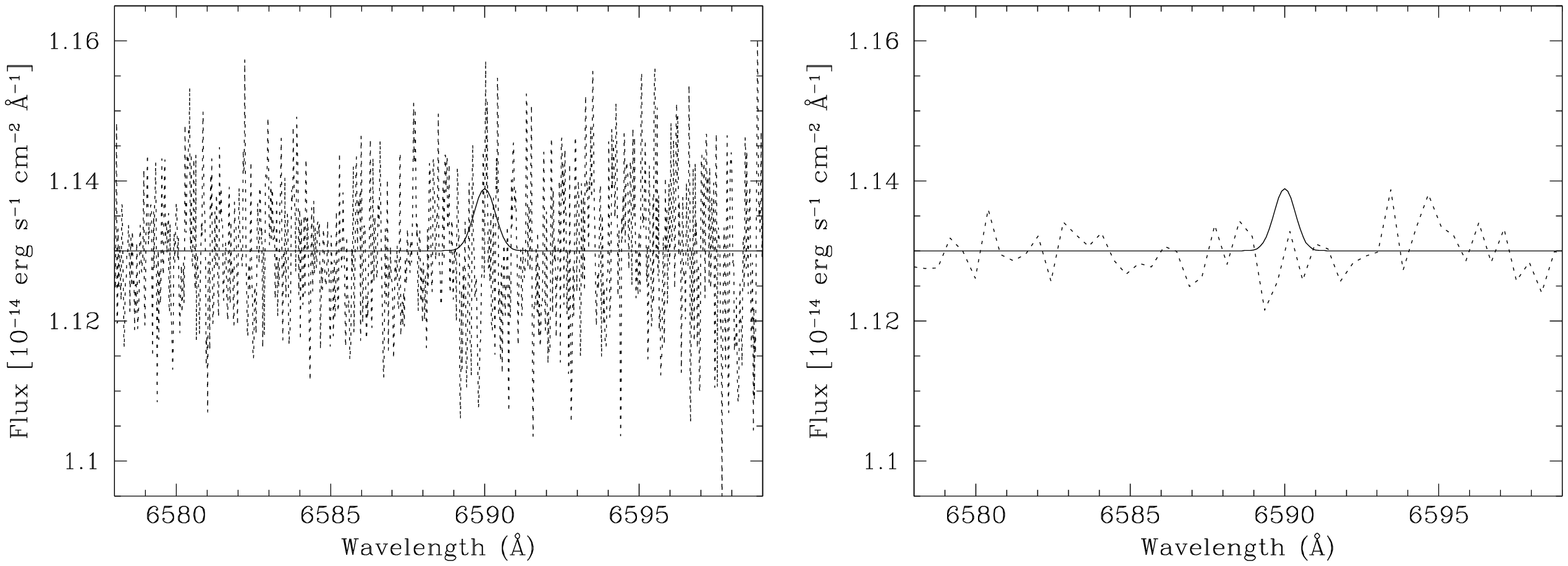}
\end{center}  
\caption{
Histograms of the line profile peak flux density distributions in the the first epoch UVES spectrum 
at the wavelength range ($\pm$2 $\times$ 190 km~s$^{-1}$) around the expected position of H$\alpha$
and assuming a line profile with FWHM of 37 km~s$^{-1}$ (upper-left) and 62 km~s$^{-1}$ 
(upper-right). The histograms have been fitted with Gaussians yielding reduced $\chi^{2}$s
of 3.2 and 1.8, respectively. The centers ($\mu$) and widths ($\sigma$) of the distributions
are given in units of 10$^{-17}$ erg s$^{-1}$ cm$^{-2}$ \AA$^{-1}$. Lower-left: The 
unbinned spectrum (0.77 km~s$^{-1}$ pixel$^{-1}$) showing the wavelength region around the expected 
position of H$\alpha$ in our first epoch spectrum. A Gaussian line profile with FWHM of 37 km~s$^{-1}$ and 
peak equal to the 3$\sigma$ level is also shown. Lower-right: The same as on 
the left except that now the spectrum has been rebinned to $\sim$18~km~s$^{-1}$ pixel$^{-1}$ resolution.}
\label{f:5}
\end{figure*}

\begin{table}
\begin{minipage}{85mm}
\caption{
Comparison of limiting fluxes from two different methods.
The center of the Gaussian noise distribution, $\mu$, and its
width, $\sigma$, yield 3$\sigma$ upper limit for the emission
line profile peak flux density, f(3$\sigma$)$_{I}$.
These can be compared to the f(3$\sigma$)$_{II}$ values derived 
simply from the measured pixel-to-pixel noise in adequately
rebinned spectra. The parameters of the flux density
distributions ($\mu$ and $\sigma$) and the upper limits
are given in units of 10$^{-17}$ erg s$^{-1}$ cm$^{-2}$ \AA$^{-1}$.
}
\begin{tabular}{lllllll}
\hline
\hline 
Line            & Epoch & FWHM          & $\mu$      & $\sigma$ & f(3$\sigma$)$_{I}$ & f(3$\sigma$)$_{II}$ \\
                &       & (km~s$^{-1}$) & \multicolumn{4}{c}{(10$^{-17}$ erg s$^{-1}$ cm$^{-2}$ \AA$^{-1}$)} \\
\hline 
H$\alpha$ $\lambda$6563$^{a}$& 1     & 37   & 0.17    & 2.9 & 8.9  & 11 \\ 
H$\alpha$ $\lambda$6563      & 1     & 62   & -0.67   & 2.6 & 7.1  & 8.1 \\ 
H$\alpha$ $\lambda$6563      & 2     & 37   & 1.1     & 3.6 & 12 & 11 \\  
H$\alpha$ $\lambda$6563      & 2     & 62   & 1.2     & 2.4 & 8.4  & 8.8 \\  
He~I $\lambda$5876$^{b}$     & 1     & 21   & 0.34    & 7.7 & 23   & 29  \\  
He~I $\lambda$5876           & 1     & 53   & 1.7     & 5.4 & 18   & 15  \\  
He~II $\lambda$4686$^{c}$    & 1     & 21   & 0.06    & 3.9 & 12   & 14 \\  
He~II $\lambda$4686          & 1     & 53   & 0.07    & 2.2 & 6.7  & 7.2 \\  
He~II $\lambda$4686          & 2     & 21   & 1.3     & 7.7 & 24   & 23  \\ 
He~II $\lambda$4686          & 2     & 53   & 1.2     & 5.0 & 16   & 16  \\ 
\hline
\hline
\end{tabular} \\
$^{a}$  The H$\alpha$ upper limits were derived for the wavelength range 6580~\AA -- 6597~\AA.\\
$^{b}$ The He~I upper limits were derived for the wavelength range 5891~\AA -- 5910~\AA. \\
$^{c}$ The He~II upper limits were derived for the wavelength range 4695~\AA -- 4713~\AA. \\
\end{minipage}
\end{table}

\begin{table}
\begin{minipage}{85mm}
\caption{3$\sigma$ upper limits for circumstellar emission line
fluxes (in units of 10$^{-17}$ erg s$^{-1}$ cm$^{-2}$) assuming 
an average temperature of T = 28~000~K (see Lundqvist et al. 2006
for details). The upper limits are shown for the wind velocities ranging
between 10 km~s$^{-1}$ and 50 km~s$^{-1}$, i.e., the resulting line widths of
FWHM = 37$-$62 km~s$^{-1}$ and 21$-$53 km~s$^{-1}$ for hydrogen and helium 
lines, respectively. For the first epoch observation the H$\alpha$
upper limit is also given for the wind velocities of 100 km~s$^{-1}$ and 200 km~s$^{-1}$.
}
\begin{tabular}{lllll}
\hline
\hline 
Line                    & $V_{\rm wind}$ & FWHM          & Epoch~1  &  Epoch~2 \\
                        & (km~s$^{-1}$)  & (km~s$^{-1}$) & \multicolumn{2}{c}{(10$^{-17}$ erg s$^{-1}$ cm$^{-2}$)} \\
\hline
H$\alpha$ $\lambda$6563 & 10             & 37            & $<$7.7      &  $<$10    \\
H$\alpha$ $\lambda$6563 & 50             & 62            & $<$10       &  $<$12   \\ 
H$\alpha$ $\lambda$6563 & 100            & 106           & $<$15$^{a}$ &  $-$\\
H$\alpha$ $\lambda$6563 & 200            & 203           & $<$20$^{a}$ &  $-$\\ \hline
He~I $\lambda$5876 & 10             & 21            & $<$10       &  $-$$^{b}$      \\
He~I $\lambda$5876      & 50             & 53            & $<$20       &  $-$$^{b}$      \\ \hline
He~II $\lambda$4686 & 10             & 21            & $<$4.2$^{c}$&  $<$8.4    \\
He~II $\lambda$4686     & 50             & 53            & $<$5.9$^{c}$&  $<$14   \\
\hline
\hline
\end{tabular} \\
$^{a}$ The upper limits for $V_{\rm wind}$ of 100 km~s$^{-1}$ and 200 km~s$^{-1}$
are based on the pixel to pixel standard deviations in the rebinned (1 pixel = FWHM/2) 
spectra between 6570~\AA~and 6608~\AA.
\\
$^{b}$ Due to unsuccessful telluric correction the He~I upper limits
for the 2nd epoch are significantly higher than for the 1st epoch observation
and therefore are not meaningful to present here.\\
$^{c}$ The on-source exposure time for the wavelength region around He~II
$\lambda$4686 was twice as long as for the regions around the other lines 
observed in the first epoch since this region was covered by both the 437 and 
564 settings of the UVES instrument.\\
\end{minipage}
\end{table}

\subsection{Early time spectra}
The overall UVES spectrum obtained on Sept. 21.3 (UT), i.e., 8.7 days before 
the B-band maximum light is shown in Fig.~\ref{f:2}. The spectrum of SN~1990N 
observed at a very early epoch of $\sim$14 days before the B-band maximum light 
(Leibundgut et al. 1991) is also shown for comparison. Although SN~1990N was
at a much earlier phase than SN~2001el at the time of the observation 
these spectra appear remarkably similar. Both SNe show strong Ca~II IR triplet and 
H\&K lines at similar high velocities. The characteristic Si~II `6150~\AA'
absorption appears flat-bottomed in both the spectra. To our knowledge such flat-bottomed Si~II lines 
have not been seen in any other SNe Ia. The origin of these lines will be discussed 
in Sect.~5.

The strongest lines expected to form in a hydrogen or helium-rich
companion wind, and which are covered by our spectral range, i.e.,
H${\alpha}$, or He~I~$\lambda$5876 and He~II~$\lambda$4686 (cf. Lundqvist
et al. 2006), were searched for in the UVES spectra at the two epochs 
of observation (see Figs.~3-4). For the predicted temperatures 
of $\sim$20~000 $-$ 40~000 K (Lundqvist et al. 2006) thermal broadening dominates the 
line widths for wind velocities up to $\sim$40 km~s$^{-1}$ for hydrogen and $\sim$20 km~s$^{-1}$ 
for helium. Here we will assume an average temperature of 28~000 K (for details see Lundqvist et al. 
2006) and a wind velocity, $V_{\rm wind}$, between 10 km~s$^{-1}$ and 50 km~s$^{-1}$, and therefore adopt observed line widths 
of 37$-$62 km~s$^{-1}$ and 21$-$53 km~s$^{-1}$ for hydrogen and helium lines, respectively. 

No narrow lines with the expected line widths were detected in our spectra and therefore only upper 
limits for the emission line fluxes were obtained. To estimate detection limits for the narrow CSM 
emission lines we used the $\tt IRAF$ FITPROFS task to perform least squares fitting of Gaussian 
profiles (one at a time) to the normalised and unbinned spectra. The free parameters were the
amplitude, the width (sigma) and the wavelength position of the Gaussian. As the spectra were already 
normalised, we did not fit the background. The 1$\sigma$ width of the
Gaussian profile was fixed according to the relevant FWHM, and the line center was moved through the 
spectrum in small (FWHM/2) steps, i.e., only the amplitude of the profile was kept as a free parameter.
To obtain statistically meaningful upper limits lines were fitted to a $\pm$2 $\times$ 190 km~s$^{-1}$ region 
around the most probable location of the CSM line with similar noise characteristics. This procedure 
yields a list of potential emission and absorption 
lines through the spectrum. In order to derive reliable flux limits from the data, the histogram distributions 
of the peak flux densities of the line profiles were inspected. These histograms have centers close to
zero flux density and can be fitted with Gaussian profiles (see Fig.~5 (upper panel)) yielding reduced $\chi^{2}$s of around
2-3 as expected from pure random noise.
The centers ($\mu$) and 1$\sigma$ widths of these distributions together with the derived 3$\sigma$ emission line
upper limits (f(3$\sigma$) = $\mu$ + 3 $\times$ $\sigma$) are listed in Table~2. For comparison we also
show 3$\sigma$ upper limits derived simply from the measured pixel-to-pixel standard deviations in the
spectra rebinned to have a Nyquist sampling, i.e., one pixel equal to FWHM/2 for each expected line width (see above).
This simple approach yields 3$\sigma$ limits consistent with the more mathematically robust method
described above.

In Fig.~\ref{f:5} (lower panel) we show the wavelength region around the expected position of the H$\alpha$ line 
in our first epoch spectrum together with a Gaussian profile with FWHM = 37 
km~s$^{-1}$ and its peak at the 3$\sigma$ level as obtained using the method described above. 
For lines of 20 $-$ 60 km~s$^{-1}$ the reduced UVES data with a 0.77 km~s$^{-1}$ pixel size 
are strongly over-sampled. Therefore, a comparison of the upper limits derived above with 
sufficiently rebinned data is more meaningful. 
In Fig.~\ref{f:5} (lower panel) we show both the unbinned (0.77 km~s$^{-1}$~pixel$^{-1}$)
spectrum and a spectrum rebinned to 18 km~s$^{-1}$~pixel$^{-1}$. We note that there are 
no features with fluxes at the level or higher than the estimated 3$\sigma$ level present in 
the rebinned data at the expected wavelength range of H$\alpha$.
The derived 3$\sigma$ upper limits for the hydrogen and helium lines at the two different 
epochs are given in Table 3. For the first epoch observation we also give H$\alpha$
upper limits for higher wind velocities of 100 km~s$^{-1}$ and 200 km~s$^{-1}$. These limits
are simply based on the pixel-to-pixel standard deviations in the rebinned (1 pixel = FWHM/2)
spectra between 6570~\AA~and 6608~\AA. 

\subsection{Late time spectra}
\begin{figure*}[t]
\begin{minipage}{180mm}
\begin{center}
\includegraphics[width=100mm, angle=90, clip] 
{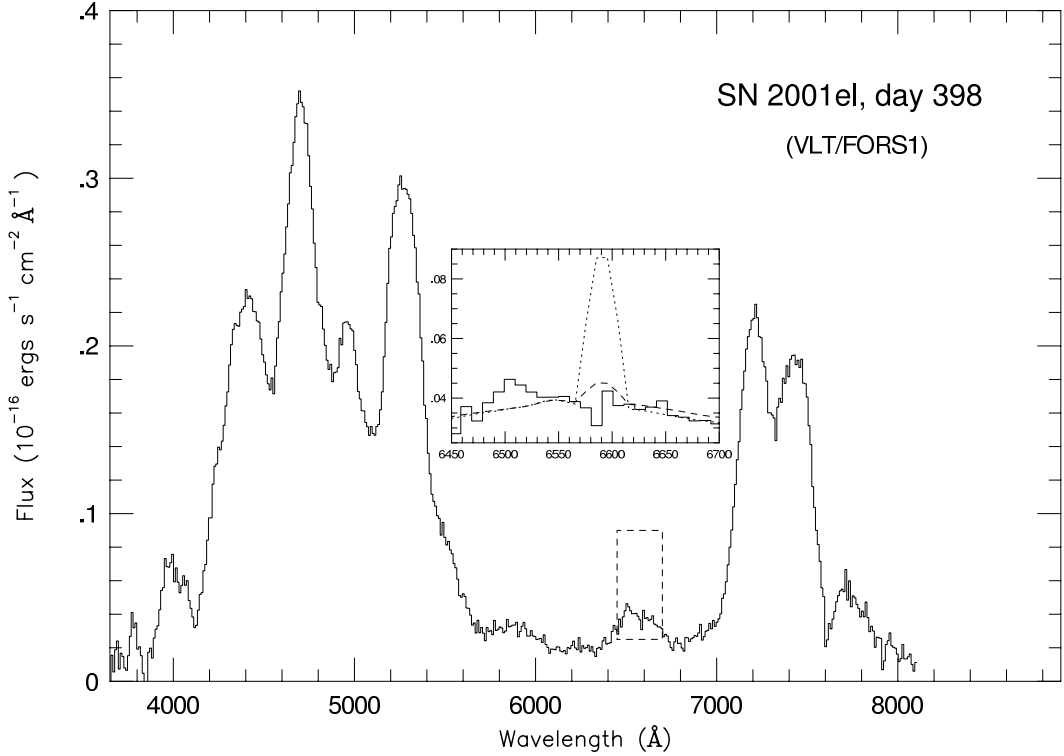}
\end{center}
\caption{
FORS1 spectrum of SN~2001el obtained on November 2.2 (UT) 2002, i.e.,
398 days past the B-band maximum light. The inset shows
an enlargement of the region marked by dashed lines, focusing on wavelengths
around H$\alpha$ for the redshift of the SN. The SN emission at these wavelengths 
is dominated by a blend of Fe~II emission lines (e.g., 6561~\AA). 
The dashed and dotted spectra in the inset are models calculated at 380 days
for the SN, assuming 0.01 \msun~and 0.05 \msun~of solar abundance material, respectively,
concentrated to $\pm 1000$ \kms~from the rest velocity of the supernova.
This is to simulate H$\alpha$ emission from gas removed from a possible
companion star according to the models of Marietta et al. (2000). 
There is no sign of such
H$\alpha$ emission in SN~2001el at these wavelengths, which places a limit
on hydrogen rich material of $\sim0.03$ \msun. The model used to calculate
the late emission is described in Sect.~4.3.
\label{f:7}
}
\end{minipage} 
\end{figure*}

In Fig.~\ref{f:7} we show the observed nebular spectrum  of SN~2001el at 398 days
after maximum light. The SN Ia spectra at these epochs are dominated by Fe
emission. The features between 4000~\AA\ and~5500 \AA\ are blends of Fe~II
and Fe~III emission, while the features in the range 7000~\AA~to 7600~\AA\
are mainly due to Fe~II and Ni~II. By comparing the observations to detailed 
late time modelling we are able to put a conservative upper limit on the amount 
of solar abundance material evaporated from a binary companion.
Following the simulations of Marietta et al. (2000) we have searched in 
our spectra for signs of relatively narrow (FWHM $\lsim$ 1000 \kms) H$\alpha$ 
emission at the range of 
wavelengths corresponding to the recession velocity found in Sect.~3.1. No such 
emission is apparent in the spectrum. Our model calculations are described in 
Sect.~4.3.

\section{Constraints on the progenitor system of SN~2001el}

\subsection{Extinction and distance of SN~2001el}
In order to estimate the upper limit for the mass loss rate, \mdot, from the 
progenitor system of SN~2001el we need to know the upper limits for the line 
luminosities, i.e., we need to know the extinction and distance towards
the supernova. Krisciunas et al. (2003) have measured a colour 
excess of $E(B-V) = 0.253\pm0.063$ for SN~2001el (see also Sollerman et al.
2005) using the Lira relation (Phillips et al. 1999). Here we adopt
$E(B-V) = 0.25\pm0.06$, i.e., A$_{V}$ = 0.78$\pm$ 0.19 (assuming R$_{V}$ = 3.1)
for the extinction towards the supernova. 

Probably the most accurate method for deriving distances for normal SNe Ia
is by making use of their photometry. Krisciunas et al. (2003) observed 
${\rm m(H)}_{14} = 13.18\pm0.05$ for the SN at +14 days from the B-band maximum 
light. Meikle (2000) estimated ${\rm M(H)}_{14} = -18.22\pm0.05$ 
for ``IR-normal'' SNe Ia. Assuming the Galactic extinction law of 
Rieke $\&$ Lebofsky (1985) we have an H-band extinction of ${\rm A}_{\rm H} = 
0.14\pm0.03$ towards the supernova. This results in a distance of $17.9\pm0.7$ Mpc 
for the supernova. We note that Krisciunas et al. (2003) obtained exactly the same distance
($17.9\pm0.8$ Mpc) for SN~2001el using their BVI light curves and the method of 
Phillips et al. (1999). As the two distance estimates obtained by making use of 
the supernova photometry are identical we adopt 17.9$\pm$0.8 Mpc for the distance of 
SN~2001el. 

\subsection{Upper limit for the progenitor mass loss rate}
We used our estimated upper limits for the narrow H$\alpha$ line luminosity 
together with the photoionisation models from Lundqvist et al. (2006) to 
obtain upper limits for the mass loss rate from the supernova progenitor 
system. In these calculations wind velocities of 10 km~s$^{-1}$ and 50 km~s$^{-1}$ and 
solar abundances were assumed. The model in Lundqvist et al. (2006) is an
updated version of that described in Cumming et al. (1996). We still
assume spherical symmetry and power-law density distributions of the SN 
ejecta and the circumstellar gas, where the ejecta profile is assumed to
follow $\rho_{\rm ejecta} \propto r^{-7}$, and the wind has the
standard $\rho_{\rm wind} \propto r^{-2}$ distribution. The shock structure,
and blast wave expansion, can then be described by the similarity
solutions of Chevalier (1982). The shocked gas produces hard radiation
that ionises the unshocked wind and gives rise to narrow circumstellar
lines. A difference compared to Cumming et al. (1996) is that we now assume 
much higher, and therefore more realistic, velocities of the ejecta
(see also Sect.~6). The velocity of the fast circumstellar shock is in our 
present models $V_{\rm s} \sim 4.45\EE{4}$ \kms~at day 1, which is about 
twice as fast as in Cumming et al. (1996). The use of such a high shock 
velocity is supported by the recent observations of high velocity Ca~II
lines in early SN Ia spectra (see Section 5 and references therein)
with their maximum velocities up to $\sim$35 000 \kms~, especially since the
circumstellar shock can be expected to advance $\sim$1.2-1.3 times (Chevalier 
1982) faster than the fastest unshocked ejecta that gives rise to the Ca~II 
lines. In addition, 
the maximum velocity of unshocked ejecta decreases with time, so the
bluest absorption observed in the ejecta 1-2 weeks before the maximum
light (see Table 3) can only give a lower limit to the maximum velocity
of unshocked ejecta at even earlier times. A dense wind will therefore 
be overtaken much faster. Compared to Cumming et al. (1996) we have also 
calculated the free-free emission from the shocked ejecta more accurately 
(cf. Lundqvist et al. 2006).

\begin{figure*}[t]
 \begin{minipage}{180mm}
\begin{center}
\includegraphics[width=150mm, angle=0, clip] {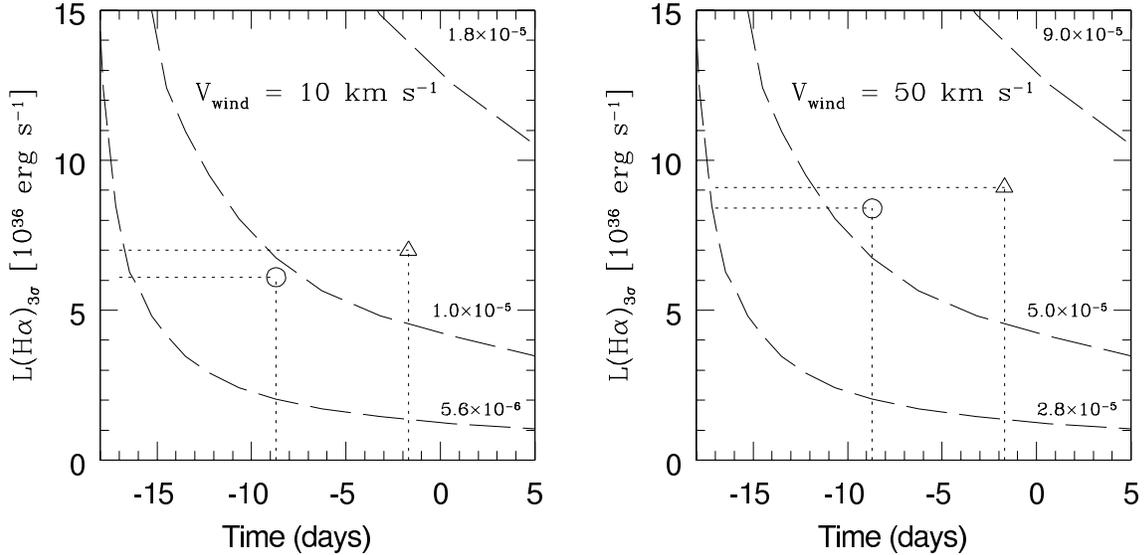}
\end{center}
\caption{
The modelled evolution of the H$\alpha$ narrow emission line luminosities
are shown for the wind velocity of 10 km~s$^{-1}$ (left) and 50 km~s$^{-1}$ (right)
as a function of the time relative to the B-band maximum light (assuming a rise time
of 19.5 days for SN~2001el). The three different 
curves (dashed lines) correspond to mass loss rates of $5.6\times10^{-6}$ M$_{\odot}$ yr$^{-1}$, 
$1.0\times10^{-5}$ M$_{\odot}$ yr$^{-1}$ and $1.8\times10^{-5}$ M$_{\odot}$ yr$^{-1}$, respectively
in the LH figure, and $2.8\times10^{-5}$ M$_{\odot}$ yr$^{-1}$, $5.0\times10^{-5}$ M$_{\odot}$ yr$^{-1}$ 
and $9.0\times10^{-5}$ M$_{\odot}$ yr$^{-1}$ in the RH figure. The 3$\sigma$ upper 
limits for the H$\alpha$ line luminosity at the two epochs of observation, $-8.7$ 
and $-1.7$ days, are marked with a circle and triangle, respectively.} 
\label{f:6}
\end{minipage}
\end{figure*}

Adopting an extinction of A$_{V}$ = 0.78$\pm$ 0.19, a distance 
of $D = 17.9\pm0.8$ Mpc, and $V_{\rm wind} = 10$ \kms~for the SN~2001el 
progenitor system, we obtain 3$\sigma$ upper limits for the H$\alpha$ narrow 
line-luminosity of $L_{\rm{H}\alpha} <5.3\times10^{36}$ erg~s$^{-1}$ and 
$L_{\rm{H}\alpha} <6.9\times10^{36}$ erg~s$^{-1}$ for the first ($-8.7$ days) 
and the second ($-1.7$ days) epoch, respectively. If a higher wind velocity of 
50 \kms~is assumed, the luminosity upper limits are $L_{\rm{H}\alpha} <6.9\times10^{36}$
erg~s$^{-1}$ and $L_{\rm{H}\alpha} <8.3\times10^{36}$ erg~s$^{-1}$ for the two epochs,
respectively.  
In Fig.~\ref{f:6} these upper limits are plotted together with modelled 
line luminosities for six different progenitor mass loss rates. Here we have 
assumed a rise time of 19.5 $\pm$ 0.2 days for SN~2001el which is typical for normal 
($\Delta$m$_{15}$(B) = 1.1) SNe Ia (Riess et al. 1999). Mass loss rate and luminosity 
are close to linearly proportional to each other in the logarithmic space (see Lundqvist et al.
2006 for details). Therefore the 
uncertainties in the rise time, distance and extinction have only a small effect in 
the derived mass loss rates. Assuming a wind velocity 
of 10 \kms~our observed line luminosity limits therefore correspond to mass loss 
rates of about $<8.9\times10^{-6}$ M$_{\odot}$ yr$^{-1}$ and $<1.2\times10^{-5}$ M$_{\odot}$ yr$^{-1}$, 
respectively. Somewhat higher upper limits of $\lsim$$1.5\times10^{-5}$ M$_{\odot}$ yr$^{-1}$ 
have previously been obtained for SNe 1994D (Cumming et al. 1996) and 2000cx (Lundqvist et al. 2006) 
also assuming $V_{\rm wind} = 10$ \kms. However, if $V_{\rm wind} = 50$ \kms~is 
assumed instead, the upper limits increase to about $<4.9\times10^{-5}$ M$_{\odot}$ yr$^{-1}$
and $<6.9\times10^{-5}$ M$_{\odot}$ yr$^{-1}$ for the two epochs, respectively.
The mass loss rates observed for symbiotic stellar systems are in the range 
between 10$^{-8}$ M$_{\odot}$ yr$^{-1}$ and $5\times10^{-5}$ M$_{\odot}$ yr$^{-1}$ (Seaquist et al.
1993) assuming $V_{\rm wind} = 30$ \kms. If adopting $V_{\rm wind} = 10$ \kms~or
50 \kms~instead, the highest mass loss rates observed for so called
Mira type systems with a red giant secondary star would be $\sim$$2\times10^{-5}$ M$_{\odot}$ yr$^{-1}$
or $\sim$$8\times10^{-5}$ M$_{\odot}$ yr$^{-1}$, respectively. Therefore, our 
results are not consistent with a symbiotic stellar system in the upper mass 
loss rate regime being the progenitor of SN~2001el. (See also Sect. 6.)

For He~I and He~II the 3$\sigma$ line luminosity upper limits are 
$L_{\rm{He}} \lsim7\times10^{36}$ erg~s$^{-1}$ and $\lsim3\times10^{36}$ erg~s$^{-1}$,
respectively, for the first epoch observation assuming a 10 \kms~wind. These are well
above the line luminosities predicted by the photoionisation models
of Lundqvist et al. (2006). However, future observations covering the He~I 
$\lambda$10830 might give more meaningful results for a helium rich 
wind.

\subsection{Upper limit for the gas lost from the companion due to the SN explosion}
Marietta et al. (2000) presented two-dimensional simulations of the
impact of a SN Ia explosion on a companion star.
They studied three different scenarios in which they assumed
the companion star to be a main-sequence star, a sub-giant
or a red giant star. They found that the amount of of hydrogen rich material 
lost varied between $\sim$0.002 \msun~and $\sim$0.5 \msun~depending
on the companion star. They also found that most of the hydrogen-rich gas 
received velocities below $\sim$1000 \kms~in all
three scenarios. In order to search for traces of hydrogen in the SN Ia ejecta as revealed 
by late time emission we have modelled nebular spectra using the code 
originally described in Kozma \& Fransson (1998), and updated 
as described in Sollerman et al. (2004) and Kozma et al. (2005). 
The calculations are based on the explosion model W7 (Nomoto et al 1984; 
Thielemann et al. 1986) where we have artificially included varying amounts 
of solar abundance material. 

Based on the calculations by Marietta et al. (2000) we fill the inner region,
out to 1000 \kms, in the W7 model with varying amounts of 
solar abundance material. Four time dependent calculations were performed 
including 0.01, 0.05, 0.10, and 0.50 \msun~of solar abundance material, 
(for further details see Lundqvist et al. 2006).
As before an extinction of A$_{V} = 0.78$ and a distance of $D = 17.9$ Mpc 
were adopted for SN~2001el. Marietta et al. (2000) also studied the asymmetry 
of the ejecta due to the impact. The asymmetry effects were not taken into 
account in our one-dimensional spherically symmetric models.
However, the asymmetries will mainly change the line profiles and will not affect 
our conclusions on the hydrogen emission.

In Fig.~\ref{f:7} we show, together with the observations of SN~2001el,
two modelled spectra at 380 days past explosion containing 0.01 \msun~and 
0.05 \msun~of solar abundance material. We see that even 
in the model containing 0.05 \msun~of hydrogen rich material, significant 
H$\alpha$ emission is present. This is clearly not seen in the observed spectrum. 
We therefore conclude that $\sim$0.03 \msun~of solar abundance material would 
have been detected in our late time spectrum via relatively narrow 
(FWHM $\lsim$ 1000 \kms) H$\alpha$ emission. Even such small amounts
of solar abundance material show up in the nebular spectrum because the optical
depth to gamma-rays in the central, hydrogen rich region is sufficiently high to 
capture enough gamma-rays to power the radiation from hydrogen. This is in 
contrast to the Fe/Ni rich lower density regions that dominate the emission in 
the spectrum in Fig.~\ref{f:7} which are predominantly powered by local positron 
deposition. The mixing of newly synthesised material into the central regions would 
increase the deposition of positrons in the hydrogen rich material. Therefore, our 
derived limit on the mass of hydrogen gas from a companion in the central 
regions of the supernova ejecta should provide a conservative upper limit.

Marietta et al. (2000) found that the amount of stripped and evaporated gas 
from the companion depends on the nature of the star. The smallest amount of gas lost was 
for the main-sequence companions at the largest binary separations of six or 
more stellar radii. They lose $\lsim$0.02 \msun~of solar abundance material 
due to the impact. However, the mass lost from the main-sequence companions 
at a smaller binary separation of $\sim$3 stellar radii, subgiant and red 
giant stars ranges from 0.15 \msun~to $\sim 0.5$ \msun, respectively. Most 
of this gas has a velocity of $< 1000$ \kms\ as the gas is evaporated from
the companion. Scenarios with such high masses are clearly not favoured by 
our results. We emphasise that the amount of gas stripped off the companion
at a high velocity is much lower than the evaporated gas. High-velocity gas
may only reveal itself in early spectra (see below).

\section{High-velocity features in SNe 2001el, 1990N, and 1984A}
Early spectroscopic observations of SNe Ia covering wavelengths longer than
$\sim 7500$~\AA\ are still scarce. Such observations have been reported to show 
the Ca~II IR triplet lines ($\lambda\lambda$ 8498, 8542, 8662) at a high velocity (HV) 
compared to the SN photospheric velocities for a few SNe Ia. This includes SN~1994D at 
$-8$ days (Hatano et al. 1999), SN~1999ee at $-7$ days (Mazzali et al. 2005a), SN~2000cx 
at $+2$ days (Thomas et al. 2004), SN~2001el at $-4$ days (Wang et al. 2003; Kasen et al. 2003), 
SN~2003du at $-5$ days (Gerardy et al. 2004), and SN~2004dt at $-7$ days (Wang et al. 2004).

Kasen et al. (2003) investigated a plausible geometry for the material giving 
rise to the observed HV Ca~II IR triplet feature in SN~2001el and its
spectropolarimetric properties (Wang et al. 2003). They favour a 
clumpy shell geometrically detached from the photospheric material, and 
suggest that the HV material could be Ca rich gas originating in either
nuclear burning in the WD or from a swept up accretion disk surrounding the 
WD.

Thomas et al. (2004) modelled both the HV IR triplet, as well as the 
HV H$\&$K feature seen in SN~2000cx at $+2$ days. They found that 3D modelling 
was needed to produce satisfactory spectral fits for both the features. They 
investigated a scenario where the HV material would be hydrogen or helium rich 
gas in the SN CSM stripped from a non-degenerate companion star of the 
progenitor WD as suggested by Marietta et al. (2000; see also Sect. 4.3). The 
models of Thomas et al. (2004) require at least $\sim$10$^{-3}$ \msun\ of 
hydrogen or helium rich material in the HV shell, which they find a factor of 
ten higher than allowed by the simulations by Marietta et al. (2000) at such
a high velocity. Due to the lack of a plausible donor mechanism to the SN CSM, 
Thomas et al. (2004) conclude that the HV material is more likely to have its 
origin in the SN explosion itself. 

Gerardy et al. (2004) further investigated the scenario of solar 
abundance material in the SN CSM. They carried out modelling where fast SN 
ejecta collide with stationary or slow moving CSM material close to the SN, 
setting up a forward and reverse shock structure. They assumed that the swept 
up CSM matter and shocked ejecta matter are mixed in the interaction forming a
dense shell of gas with a suitable Ca abundance (solar abundance) to give 
rise to the observed HV lines. In their modelling three different scenarios 
produced a shell of 0.02 \msun\ at 20 days post-explosion which 
is capable of giving rise to the observed HV Ca~II IR triplet feature.
We note that the progenitor mass loss rate required by their first 
scenario is a factor of $\sim$20 higher than our upper limit for the 
progenitor of SN~2001el (assuming a 10 \kms~wind). This scenario
as well as a scenario with a constant density CSM environment were also
found unrealistic by Gerardy el al. (2004) as their energy generation
rates would clearly dominate the bolometric light curve of the SN.
For SN 2003du Gerardy
el al. (2004) favour a scenario where the matter involved in the
interaction originates much closer to the progenitor system, in an
accretion disk, Roche-lobe, or in a common envelope.

More recently, Mazzali et al. (2005a) found that the observed HV Ca~II
IR triplet and Si~II features in SN~1999ee could be explained by
an increase of the mass at the highest ejecta velocities. According to the
authors such a density increase could either be due to asymmetries in the 
explosion, and/or interaction of the ejecta with hydrogen rich material
in the CSM of the SN. They found that about 4$\times$10$^{-3}$~\msun\ of hydrogen rich 
material in the CSM could explain the HV features observed in SN~1999ee.

\subsection{High-velocity Ca~II}
In Fig.~8 we show Ca~II profiles of SN~2001el together with the profiles of two other 
SNe Ia, SN~1984A (Wegner \& McMahan 1987) and SN~1990N (Leibundgut et al. 1991)
both showing very high velocities at a similar early epoch.
For these figures the spectra were normalised by a first order polynomial 
fit around the profile of interest, and converted to the velocities 
w.r.t. to the rest wavelengths assuming $V_{\rm rec}$ = 1180 km~s$^{-1}$, 1010 km~s$^{-1}$ and -261 km~s$^{-1}$, 
for SNe 2001el, 1990N, and 1984A, respectively. To illustrate 
the maximum extent of the absorbing gas (in velocity space) we used the rest wavelengths of the bluest components of 
the Ca~II triplet/doublet lines viz. 8498~\AA~and 3934~\AA. However, due to limited wavelength coverage in the 
SN~1990N spectrum the normalisation of its Ca~II IR triplet profile is unreliable in the red part. Therefore, in the discussion 
below we will only make use of the blue part of this profile. In Table 4, we list the maximum velocities 
as indicated by the blue edges of the normalised HV profiles.

Our first epoch ($-9$ days) observation of SN~2001el constitutes to our knowledge
the earliest reported coverage\footnote{
While this paper was being refereed other studies presenting data at even earlier
epochs have appeared (see Mazzali et al. 2005b; Quimby et al. 2005).}
of the HV Ca~II IR triplet in a SN Ia (in addition to SN~1990N). 
Our $-9$ day observation covers also the Ca~II H$\&$K lines enabling comparisons between the 
velocity profiles of these two features. 
In Fig.~8a we show profiles of the HV Ca~II triplet and doublet lines as observed at $-9$ days. We 
illustrate the spectral evolution of the Ca~II IR triplet between $-9$ and $-4$ days in Fig.~8b using 
the low-resolution spectrum obtained by Wang et al. (2003) for the latter epoch. At both epochs, the IR 
triplet profile of SN~2001el shows a double dipped structure  
whereas the H$\&$K doublet (3934, 3968~\AA)
feature ($\sim$2600 \kms~separation) does not show evidence for two
components. The bluemost minimum in the IR triplet profile has been
identified to be due to the blend of the $\lambda$8542 and $\lambda$8498
components ($\sim$1500 \kms~separation) whereas the redmost minimum is
produced by the $\lambda$8662 component (Kasen et al. 2003).

The velocities of the centers of the two Ca~II profiles at $-9$ days are quite similar. However, the 
maximum velocity, as indicated by the blue edge of the profile, is higher for the Ca~II H$\&$K lines 
($\sim$34~000 km~s$^{-1}$) than for the IR triplet ($\sim$28~000 km~s$^{-1}$). Similar behaviour was
also shown by SN~1990N at $-14$ days with maximum velocities of 36~000 km~s$^{-1}$ and $\sim$28~000 km~s$^{-1}$
for the Ca~II H$\&$K and the IR triplet profiles, respectively. This behaviour can be due to blending 
of the Ca~II H$\&$K profile with other lines as already pointed out by Kasen et al. (2003) for SN~2001el, 
but may also just reflect a higher absorption in the H$\&$K lines. This can be expected since the 
H$\&$K lines arise from a lower level which may well have a higher population than the level from 
which the Ca~II IR lines are absorbed. However, estimating the non-LTE level populations and 
the exact contribution of other lines to the observed profile needs detailed modelling of the SN 
spectrum. The highest velocity of the three SNe as measured
from the blue edges of their Ca~II H$\&$K profiles is shown by SN~1984A (38~000 km~s$^{-1}$) at $-7$ days
(see Table 4).

\begin{figure*}[t]
 \begin{minipage}{180mm}
\includegraphics[width=60mm, angle=0, clip] {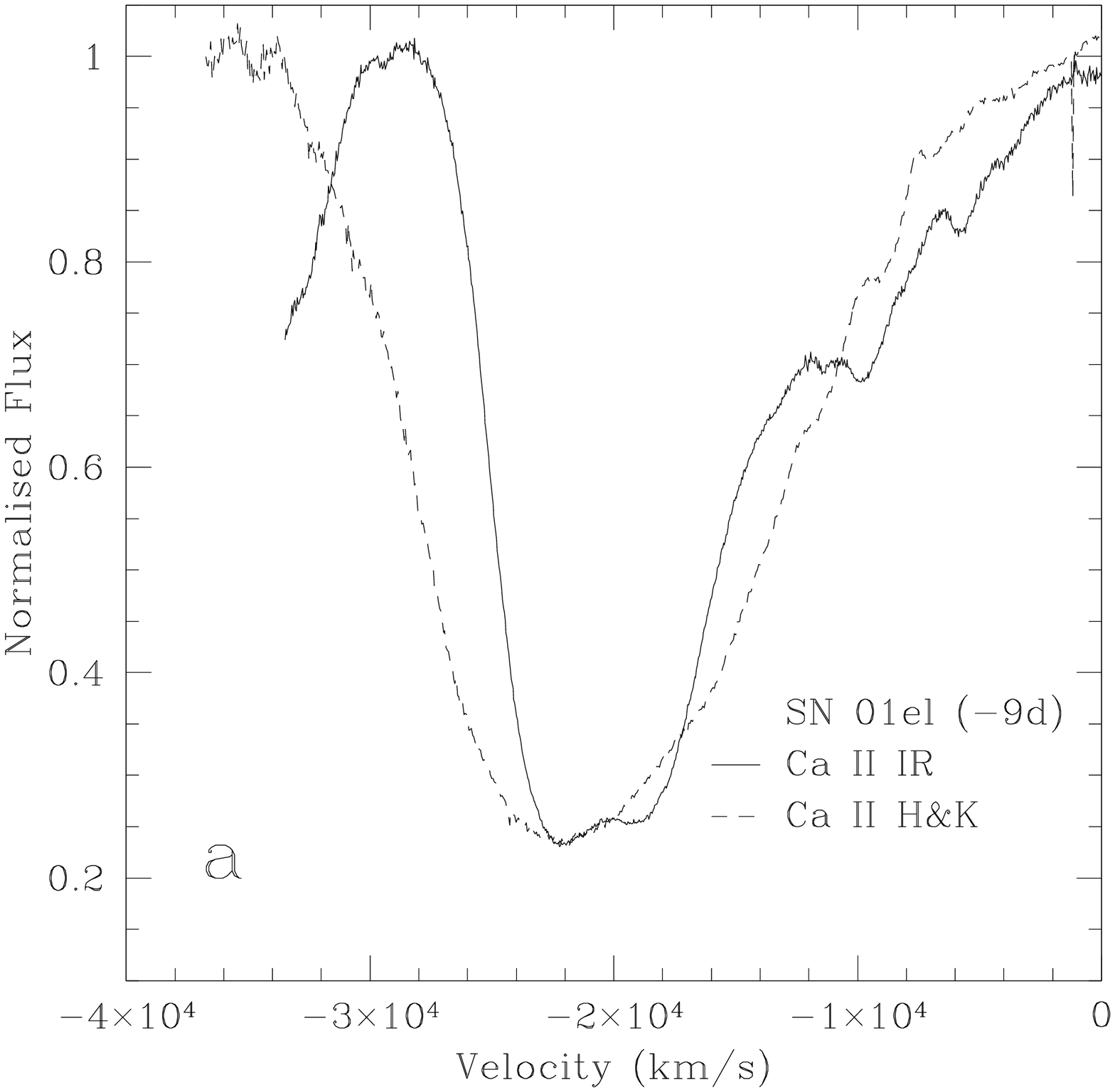}
\includegraphics[width=60mm, angle=0, clip] {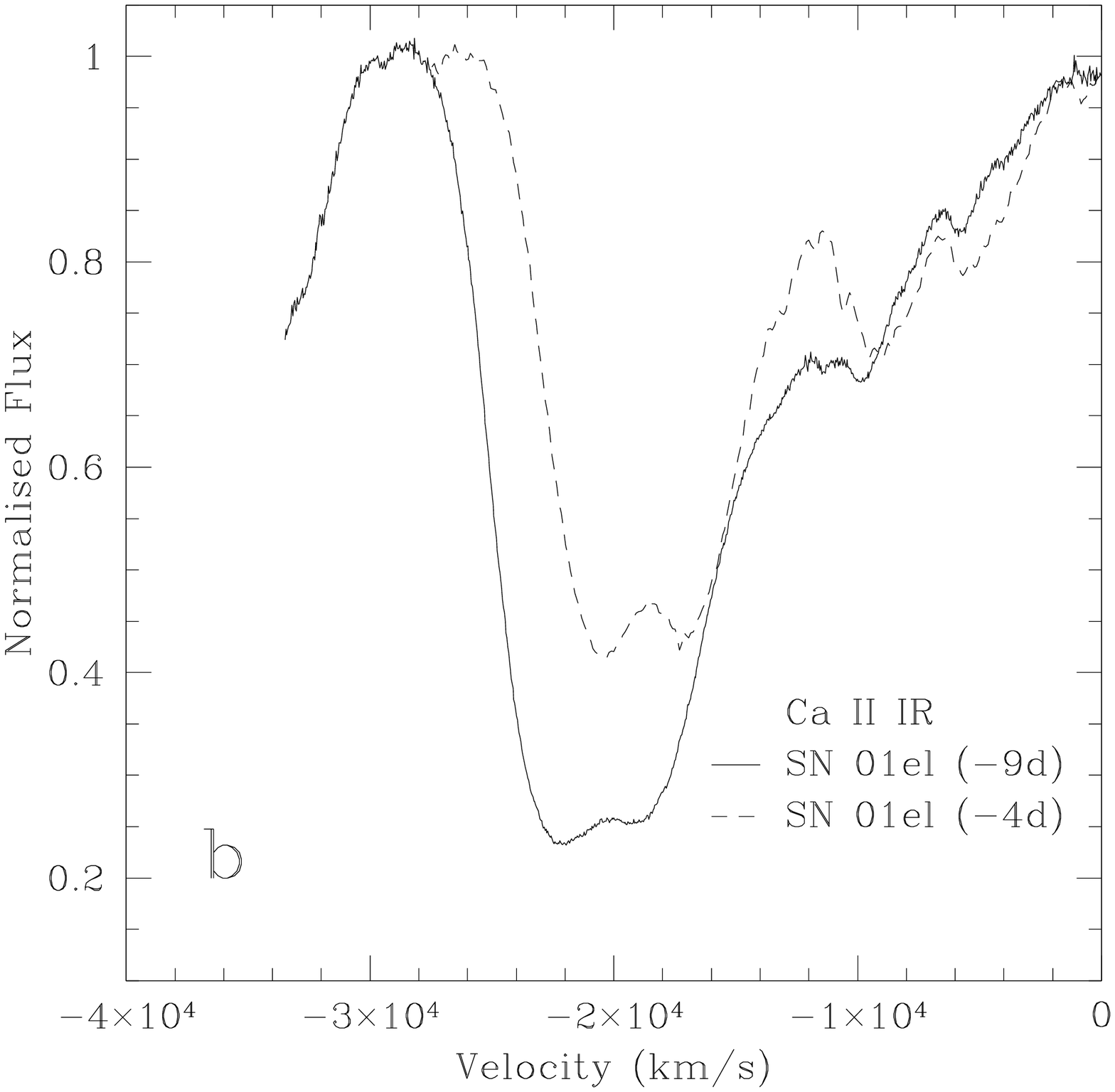}
\includegraphics[width=60mm, angle=0, clip] {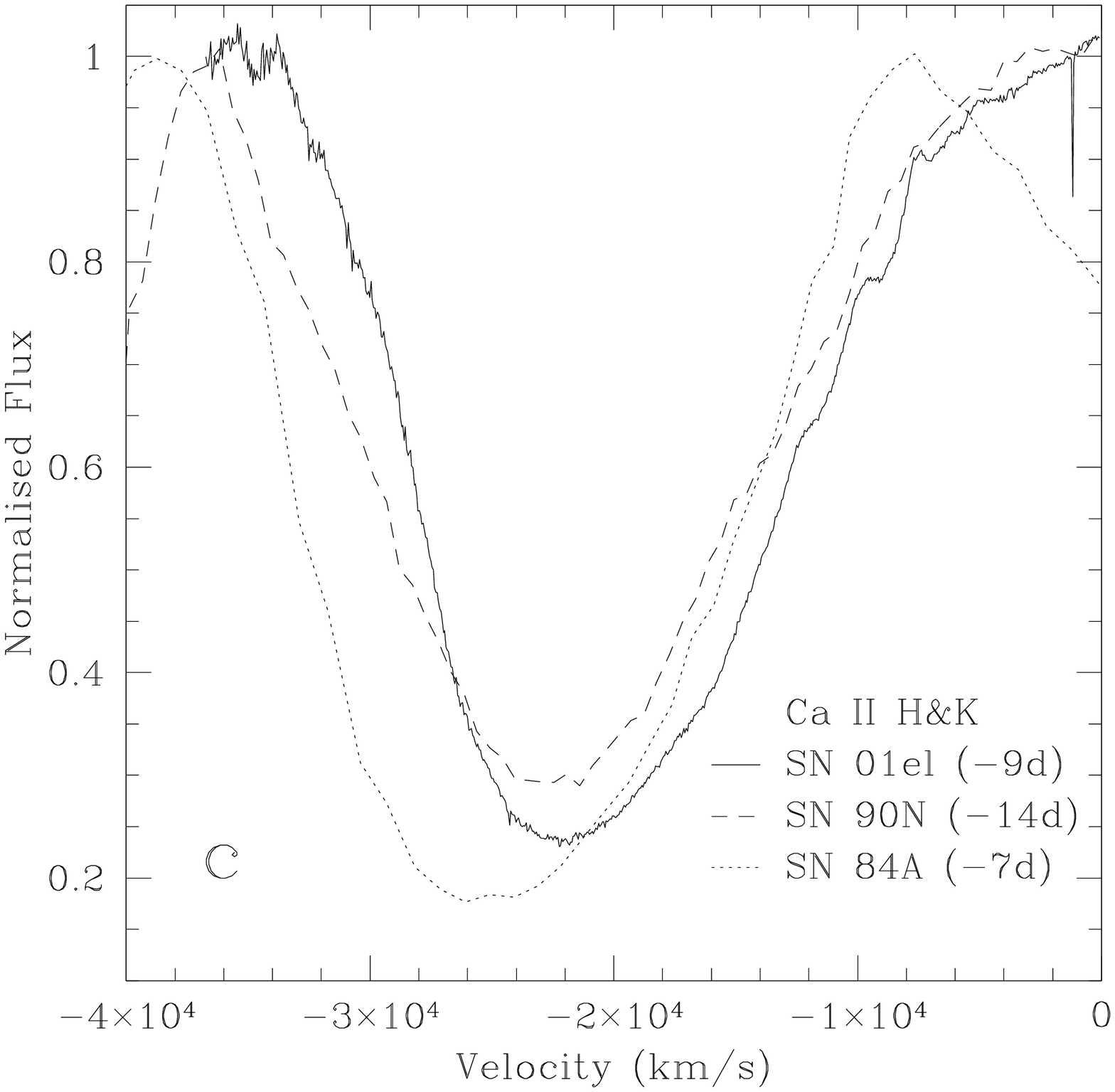}
\includegraphics[width=60mm, angle=0, clip] {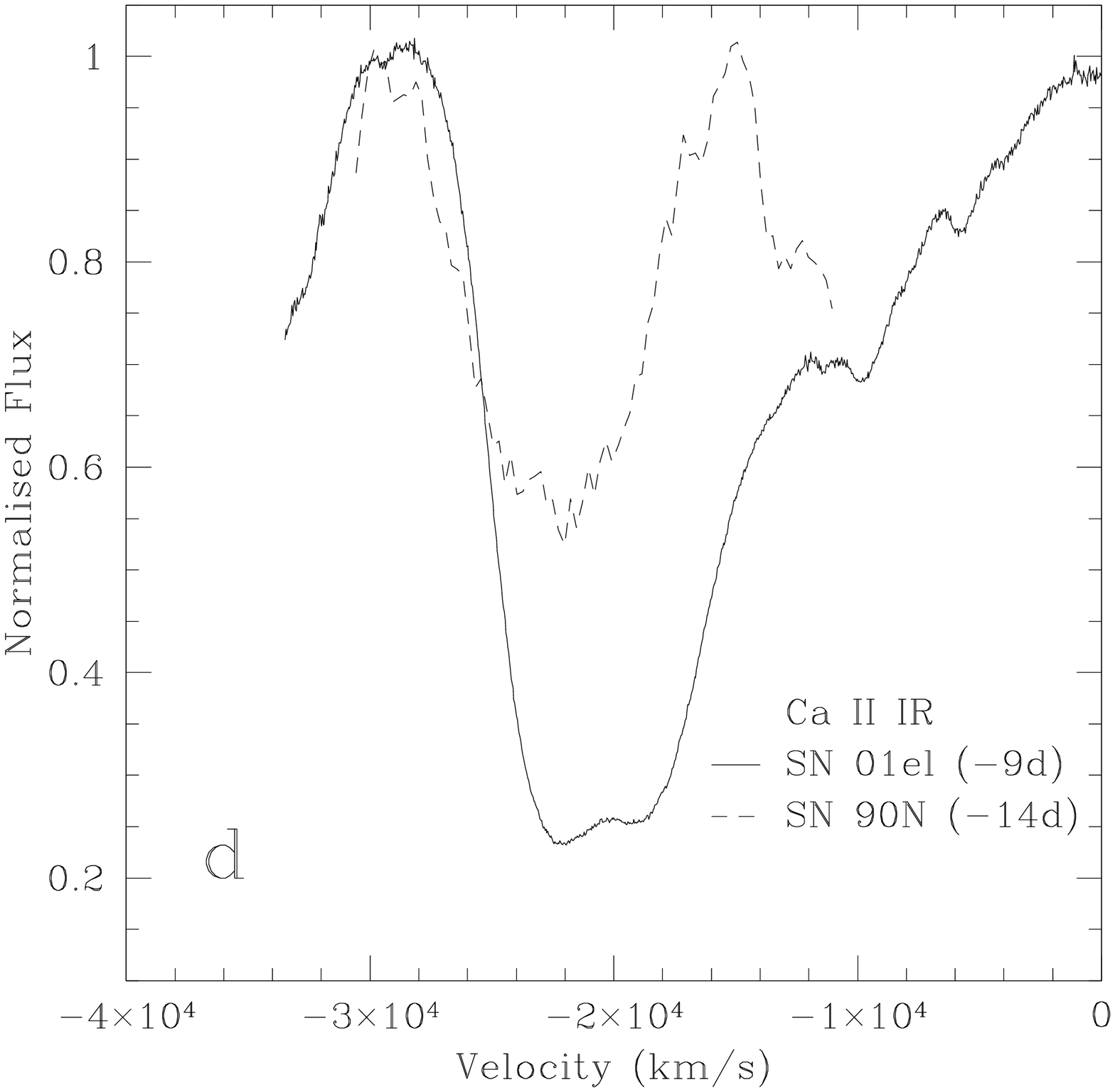}
\includegraphics[width=60mm, angle=0, clip] {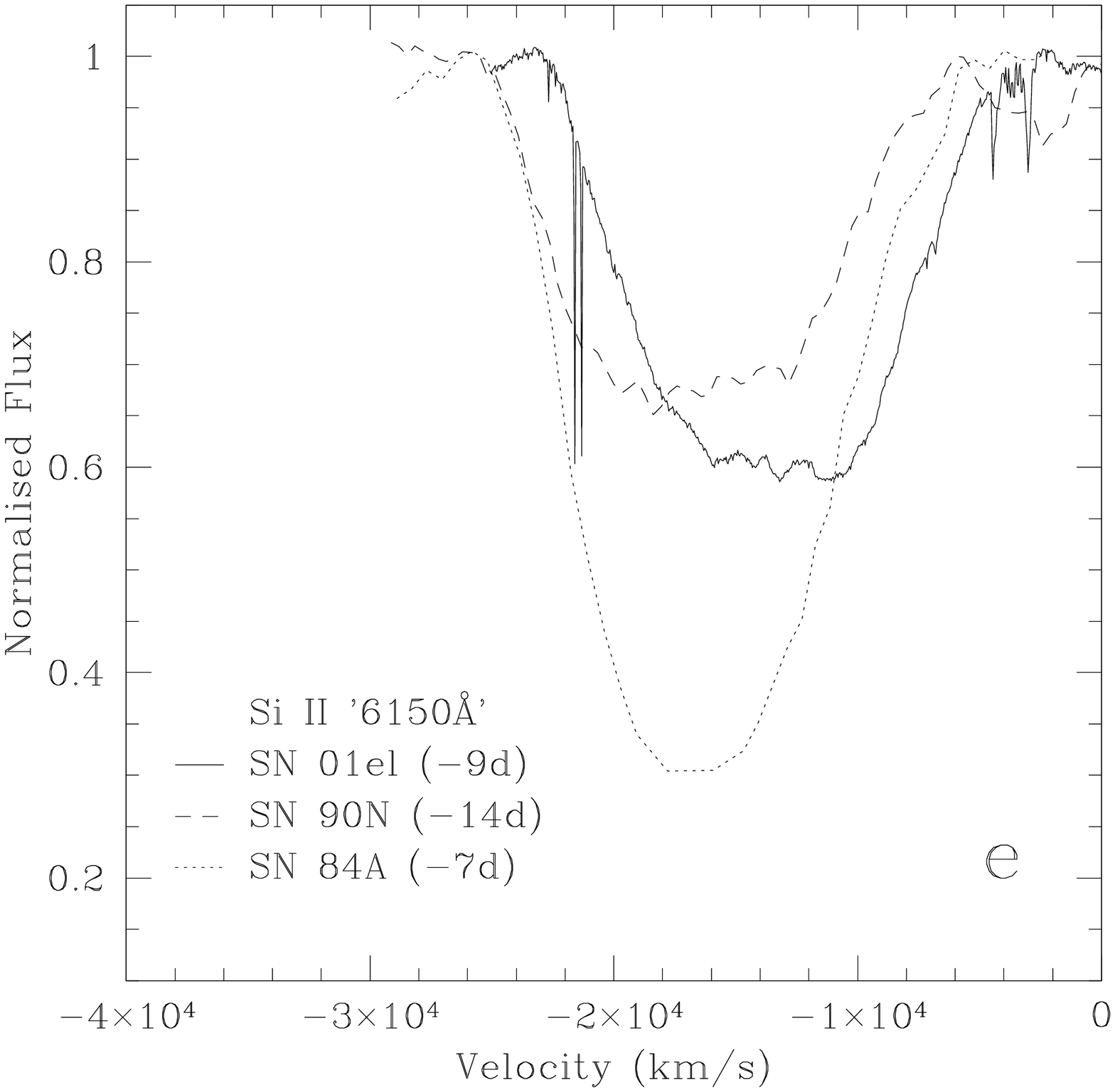}
\includegraphics[width=60mm, angle=0, clip] {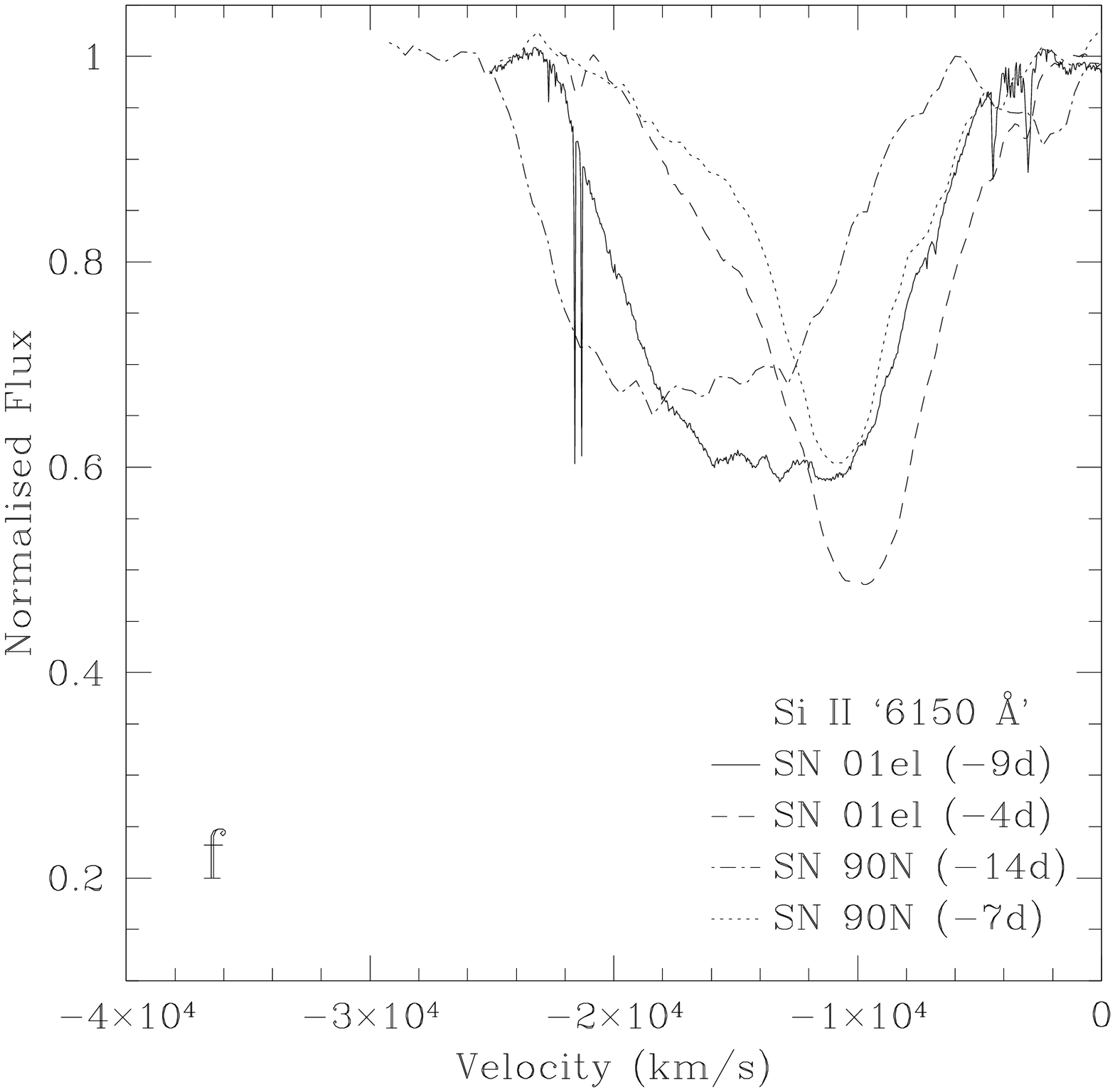}
\caption{
(a) Comparison between the Ca~II IR triplet (solid line) and
the Ca~II H$\&$K velocity profiles (dashed line) at $-9$ days. 
(b) Spectral evolution of the Ca~II IR triplet from $-9$ days 
(solid line) to $-4$ days (dashed line). 
(c) The Ca~II H$\&$K profiles in SNe 2001el at $-9$ days
(solid line), 1984A at $-7$ days (dotted line), and SN~1990N at $-14$ days
(dashed line).
(d) Comparison between the Ca~II IR triplet profiles of SNe 2001el 
(solid line) and 1990N (dashed line). Note that due to the limited wavelength
coverage in SN~1990N spectrum the normalisation of its Ca~II IR profile is unreliable
in the red part.
(e) Comparisons between the Si~II `6150~\AA' profiles in SNe 2001el at $-9$ days
(solid line), 1984A at $-7$ days (dotted line), and SN~1990N at $-14$ days
(dashed line). 
(f) Spectral evolution of the Si~II `6150~\AA' profile in SNe 2001el
(solid and dashed lines) and 1990N (dotted and dotted-dashed lines).
The flux calibrated ($f_{\lambda}$) spectral features have been normalised and 
converted to velocity space w.r.t. to the rest wavelengths assuming $V_{rec}$ = 
1180, 1010, and -261 km~s$^{-1}$ for SNe 2001el, 1990N, and 1984A, respectively. For 
this we used the rest wavelengths of the bluest 
components of the Ca~II triplet/doublet lines viz. 8498~\AA~and 3934~\AA, and
the Si~II `6150~\AA' profile of 6347~\AA.} 
\label{f:8}
\end{minipage}
\end{figure*}

\subsection{High-velocity Si~II}
In Fig.~8, the Si~II `6150 \AA' profile is shown for SNe 2001el, 1990N,
and 1984A. Again, the data were normalised and converted to velocities
w.r.t. the bluest component of the Si~II profile of 6347~\AA~
(see Table 4). In Fig.~8e, the Si~II profile of SN~2001el at 
$-9$ days is compared with the Si~II profiles of SNe 1984A and 1990N at $-7$ and
$-14$ days, respectively. At these epochs both SNe 2001el and 1990N show
a flat-bottomed Si~II profile. To our knowledge such flat-bottomed Si~II 
lines have not been seen in any other SNe Ia so far.

The pre-maximum light evolution of the Si~II profile is illustrated in 
Fig.~8f for SNe 2001el and 1990N. For the latter epochs we used spectra presented
by Wang et al. (2003) and Leibundgut et al. (1991), respectively.
In SN~2001el the flat bottom of the line disappeared by the next earliest epoch of observation, 4 
days before the maximum light. The same behaviour is also apparent in SN~1990N where the shape of the 
line has changed by the $-7$ day epoch. In Fig.~5a of Fransson (1984) such a 
flat-bottomed profile, and its evolution are presented as a result of calculations assuming that 
the velocity in the SN ejecta is proportional to its radius as expected soon after the 
SN shock outbreak. The flat-bottomed line shape is produced by pure scattering coming from 
a thin region compared to the size of the SN photosphere. As the scattering region 
widens when the SN `photosphere' receeds the flat-bottomed line shape disappears, 
becoming an ordinary P-Cygni profile (cf., the change from the $R_2=1.2$ case to the $R_2=1.5$ case 
in Fig.~5a of Fransson 1984).

The fact that the Ca~II lines extend to considerably higher
velocities than the Si~II line (see Table 4) explains the less
pronounced flat section of the Ca~II lines. The lower
maximum velocity of the Si~II line compared to the Ca~II lines is
likely to be a result of the larger population of the atomic levels of
the Ca~II lines compared to the excited Si~II line, and does not need to indicate different
abundance distributions. We note that the flat-bottomed Si~II profile in 
SN~1990N has also been explained by a blend of the Si~II line with C~II 6580~\AA~line forming in a 
carbon-rich HV shell detached from the SN photosphere (Fisher et al. 1996), but see
Mazzali (2001). Our explanation for the flat-bottomed Si~II profile
therefore gives less support to an enhanced HV carbon shell.

\begin{table}
\begin{minipage}{85mm}
\caption{
Maximum velocities for SNe 1984A, 1990N, and 2001el as indicated by the blue edges of the 
Ca~II H\&K and IR triplet and Si~II `6150~\AA' profiles (see Fig.~8).
}
\begin{tabular}{lllll}
\hline
\hline 
SN     & Epoch    & V$_{\rm Ca~II,H\&K}$ & V$_{\rm Ca~II,IR}$ & V$_{\rm Si~II}$ \\
       &(days)    & (\kms)           & (\kms)                 & (\kms) \\ \hline
1984A  &$-7$$^{a}$  & 38~000           & -                      & 26~000 \\
1990N  &$-14$$^{b}$ & 36~000           & $\sim$28~000           & 26~000 \\
2001el &$-9$        & 34~000           & 28~000                 & 23~000 \\
\hline
\hline
\end{tabular} \\
$^{a}$ Date of observation 1984 Jan 10 (Wegner \& McMahan (1987). 
The epoch adopting Jan 17 for the B-band maximum light (Barbon et al. 1989).\\
$^{b}$ Date of observation 1990 June 26.2 UT (Leibundgut et al. 1991).\\
\end{minipage}
\end{table}


\section{Discussion and Summary}
We have looked for narrow hydrogen and helium lines in our
early high-resolution optical spectra of the normal type Ia supernova 
SN~2001el. No such lines were detected, and we derived upper limits for 
the luminosities of H$\alpha$, He~I~$\lambda$5876 and He~II~$\lambda$4686. 
These luminosities were then compared with the photoionisation models of 
Lundqvist et al. (2006). From the limits for H$\alpha$ we conclude that 
the mass loss rate from the progenitor system of SN~2001el was less 
than $\lsim9\times10^{-6} \Msunyr$ for a spherically symmetric wind 
with a velocity of 10 \kms~or less than $\lsim5\times10^{-5} \Msunyr$ for 
a velocity of 50 \kms. These estimates are only sensitive to gas outside a 
few $\times 10^{15}$ cm from the SN explosion site. The gas inside this radius
would have been swept up by the SN ejecta prior to our observations.
However, these are the best H$\alpha$ based upper limits obtained for a 
SN Ia so far. Our results, together with the previous observational 
work on SN~1994D (Cumming et al. 1996) and SN~2000cx (Lundqvist et al. 2006) 
do not favour a symbiotic star in the upper mass loss rate regime (so called 
Mira type systems) from being the likely progenitor scenario for these SNe. 
This is in accordance with the models of Hachisu et al. (1999a, 1999b), 
which do not indicate a mass loss rate of the binary companion in the 
symbiotic scenario higher than $\sim$10$^{-6}$ $\Msunyr$. Stronger mass loss 
would initiate a powerful (although dilute) wind from the white dwarf that 
would clear the surroundings of the white dwarf, and that may even strip off 
some of the envelope of the companion. The white dwarf wind with its possible 
stripping effect, in combination with the orbital motion of the stars, is 
likely to create an asymmetric CSM. Effects of a more disk-like structure of 
the denser parts of the CSM, and how this connects to our spherically symmetric 
models, are discussed by Cumming et al. (1996). In general, lower values
of \mdot $/V_{\rm wind}$ should be possible to trace in an asymmetric scenario,
although uncertainties due the inclination angle and the flatness of the
dense part of the CSM would also be introduced. However, to push the
upper limits down to the $\sim$ 10$^{-6}$ $\Msunyr$ level a much more nearby 
($D \sim 3$ Mpc) SN Ia needs to be observed even earlier ($\sim 15$ days 
before the maximum light) than our first epoch observation of SN~2001el 
(Mattila et al. in prep.). This will only be feasible by making use of 
target-of-opportunity (ToO) time offered by observatories such as ESO and 
Subaru. Similar, or even lower limits on \mdot $/V_{\rm wind}$ can be obtained 
from radio and X-ray observations (e.g., Eck et al. 2002; Lundqvist et al. 
2006), but such observations cannot disentangle the elemental composition of 
the gas in the same direct way as optical observations can do.

SN~2001el was revisited using VLT/FORS in the nebular phase, $\sim$400 days 
after the maximum light. We modelled the late time spectrum to derive an upper 
limit of $\sim$0.03 M$_{\odot}$ for solar abundance material present 
at velocities lower than 1000 \kms~within the supernova explosion site. This 
is similar to our results for SNe 1998u and 2000cx (Lundqvist et al. 2006). 
Comparing this limit to the numerical simulations of Marietta et al. (2000) 
indicates that symbiotic systems with a subgiant, red giant or a main-sequence 
secondary star at a small binary separation are not likely progenitor scenarios 
for these SNe (see also below). These results demonstrate that the combination
of very early and late time spectroscopy can be a powerful tool to probe
SN Ia progenitors.

Our $-9$ day spectrum shows that both the Ca~II IR triplet and the H$\&$K lines 
are present at very high velocities of up to $\sim$28~000 km~s$^{-1}$ and
$\sim$34~000 km~s$^{-1}$, respectively. 
High-velocity Ca~II in SNe~Ia is, however, not unique for SN~2001el.
Here we have highlighted this for SNe 1984A and 1990N, which actually 
showed $\sim 2000-4000$ km~s$^{-1}$ higher velocities for the Ca~II H\&K as
well as the Si~II lines at a similar epoch. As a matter of fact, high-velocity 
Ca~II has now been seen in all SNe Ia (Mazzali et al. 2005b) which have 
been studied early enough. The presence of the lines confirm the assumption
in Lundqvist et al. (2006) that at least some SNe~Ia launch their outermost 
ejecta at velocities in excess of $c/10$. This means that they sweep 
up any CSM at a much faster rate than was assumed in the earlier models of 
Cumming et al. (1996) (see Sect.~4.2).

A fundamental question is of course the origin of the HV Ca~II H\&K lines. 
One possibility mentioned in Sect. 5 is that they form in
material stripped off from a non-degenerate companion. From our FORS 
observation at 400 days post-maximum we were able to put an upper limit 
of $\sim$0.03 \msun\ for solar abundance material with a low 
velocity ($\sim$1000 km~s$^{-1}$) evaporated from the companion star. 
According to the models of Marietta et al. (2000) only a small fraction of 
the gas lost from the companion, $<$1/1000, can have a high velocity 
of $>$15~000 km~s$^{-1}$. The mass of the solar abundance HV gas (originating 
from the companion star) available to give rise to the HV Ca~II lines would 
therefore be less than $\sim$3 $\times$ 10$^{-5}$ \msun\ which is a factor 
of $\sim$3000 
less than the mass of solar abundance gas estimated by Wang et al. (2003) 
to be required to produce the HV Ca~II IR triplet feature seen in SN~2001el. 
Our observations therefore do not support that the HV Ca~II lines would 
arise in gas stripped from the companion star, which is consistent with the
findings of Thomas et al. (2004) for SN~2000cx.

We also note that there is no H$\beta$ absorption apparent in our 
spectrum (see Fig.~\ref{f:2}) at the velocity of the HV Ca~II features. A very 
tentative identification of such a HV H$\beta$ line was made by Branch et al. 
(2004) in SN~2000cx spectra. The existence of a HV H$\beta$ line would have 
indicated a non-degenerate companion star origin for the HV matter, which is 
thus not supported by our observations.
A very low hydrogen mass upper limit of $\sim$3 $\times$ 10$^{-4}$ \msun~was 
presented by Della Valle et al. (1996) for SN~1990M based on a non-detection of 
atmospheric H$\alpha$ absorption in their spectrum near the maximum light. 
However, this was obtained by using a simplified model. H$\alpha$ is usually 
in net emission in supernovae, and thus the absorption is weaker than would 
be predicted by an analysis using resonant scattering. The emission falls 
into the Si~II absorption trough and so could be hidden in SNe Ia 
(Thomas et al., 2004; Branch et al. 2004). This effect is somewhat less for 
H$\beta$ and H$\gamma$. Lentz et al. (2002) performed detailed spectral 
synthesis calculations of mixed solar material into a W7 model and found that 
even relatively large amounts of hydrogen ($\sim$ 10$^{-3}$ \msun) would be 
hard to detect at the earliest times. Furthermore, early time observations 
can only detect hydrogen present in the outer parts of the supernova ejecta, 
or in a detached wind (see Sect.~4.2). However, as the simulations of Marietta 
et al. (2000) predict only a very small high-velocity tail for the hydrogen 
rich material, early observations seem futile to reveal any stripped material 
from the companion. The models of Marietta et al. (2000), in combination with
our results in Sect. 4.3 and in Lundqvist et al. (2006), show that 
stripped, or rather evaporated material from a companion, should instead be 
looked for at late times. Our limits on the late emission are actually so 
tight that most of the scenarios in Marietta et al. (2000) appear to be ruled 
out. This, in combination with the properties of the presumed binary companion 
of the Tycho progenitor (Ruiz-Lapuente et al. 2004), should be used to find a 
viable solution to the single-degenerate progenitor scenario.

Another possibility for the HV Ca~II lines mentioned in Sect. 5 is that they
arise in gas compressed due to circumstellar interaction (Gerardy et al. 2004). 
The scenario with a dense continuous wind outside a few$\times 10^{15}$ cm 
is not supported by our observations and models of narrow circumstellar 
emission lines. Furthermore, the lack of interaction signatures in the 
bolometric light curves (e.g. Gerardy el al. 2004) and the lack of radio emission in
normal SNe Ia (e.g., Eck et al. 2002), do not support such a scenario
either. The model with a circumstellar disk could be more likely,
although we note that such a scenario should produce a range of velocities
for the Ca~II absorption depending on the viewing angle, unless the disk
is thick. More self-consistent modelling of the circumstellar interaction
is needed to see whether a circumstellar disk model can explain the HV Ca~II
features seen in SNe Ia.

A more natural explanation for the HV Ca~II lines and the HV Si~II lines in 
SN~2001el and other SNe Ia would appear to be absorption in the outermost 
parts of the SN ejecta which extend to high enough velocities. 
We discussed this qualitatively in Sect.~5.2. Maximum 
velocities of up to $\sim$26~000 km~s$^{-1}$ and $\sim$38~000 km~s$^{-1}$ were 
observed for the blue edges of the Si~II and Ca~II H$\&$K profiles, 
respectively, in the $-7$ day spectrum of SN~1984A. Lentz et al. (2001) 
modelled the early spectra of SN~1984A and found that the large blueshifts 
could be produced by a higher density in the outer layers of the ejecta. 
Furthermore, Hatano et al. (2000) showed that even at 10 days past maximum 
light, i.e., much later than the times studied here, there exists a class 
of `fast' supernovae typified by SN~1984A. More recently Benetti et al. (2005) 
divided `normal' SNe Ia into groups showing high (HVG) and low (LVG) temporal 
velocity gradients for the Si~II profile after the maximum light. Both 
SNe 2001el and 1990N belong to the LVG group whereas SN~1984A belongs to the 
HVG SNe which also tend to show the highest photospheric velocities both 
before and after the maximum light. It is important to understand whether 
the density structure variations of the HV SNe Ia are due to initial 
conditions, variations in the explosion mechanism or something else. How 
these HV SNe Ia are correlated with brightness at maximum light is therefore 
an important question for the use of SNe Ia as cosmological probes.

\begin{acknowledgements}
The VLT/UVES observations were obtained in ToO service mode. We thank the 
Paranal staff for their help with the observations, and Robert Cumming 
for helpful discussions and comments on the manuscript. We thank Lifan Wang 
for letting us to use his spectrum of SN~2001el from the epoch $-4$ days and 
Stefano Benetti for the early spectra of SN~2002bo. S.M. acknowledges 
financial support from the `The Physics of Type Ia SNe' Research Training 
Network under contract HPRN-CT-2002-00303. This work was also supported by 
the Swedish Research Council, the Swedish Space Board, and the Royal Swedish
Academy of Sciences. P.L. is a Research Fellow at the Royal Swedish Academy 
supported by a grant from the Wallenberg Foundation. E.B. was supported in part 
by grants from NASA and the NSF. K.N. was supported by the Japan Society for 
the Promotion of Science. 
\end{acknowledgements}

\end{document}